\documentclass[usenames,dvipsnames]{article}

\usepackage{a4wide}
\usepackage{bbm}
\usepackage{amsmath,amsfonts,amsthm,amssymb}
\usepackage[justification=justified, singlelinecheck=false]{caption}
\usepackage{graphicx}
\usepackage{enumitem}

\usepackage{tikz}

\usepackage{algorithmicx}
\usepackage{algorithm}
\usepackage{algpseudocode}

\usepackage{wasysym} 
\usepackage{here}
\usepackage{hyperref}
\usepackage{ulem}
\usepackage{mwe}

\usepackage[scaled=.90]{helvet}
\usepackage{courier}
\usepackage{ae}
\usepackage[T1]{fontenc}
\usepackage{subfigure}
\usepackage{graphicx,color}
\usepackage{xcolor}

\usepackage[T1]{fontenc}
\usepackage{here} 
\usepackage{xpatch}
\makeatletter
\xpatchcmd{\@thm}{\thm@headpunct{.}}{\thm@headpunct{}}{}{}
\makeatother

\newcommand{\norm}[1]{\left\lVert#1\right\rVert}

\newtheorem*{notation}{Notation}
\newtheorem{remark}{Remark}[section]

\theoremstyle{definition}

\begin{document}
	
	\title{Network inference via approximate Bayesian computation. Illustration on a stochastic multi-population neural~mass~model}
	\author{Susanne Ditlevsen\footnotemark[1]\thanks{Department of Mathematical Sciences, University of Copenhagen (susanne@math.ku.dk)}, Massimiliano Tamborrino\footnotemark[2]\thanks{Department of Statistics, University of Warwick (massimiliano.tamborrino@warwick.ac.uk)}, Irene Tubikanec\footnotemark[3]\thanks{Institute of Applied Statistics, Johannes Kepler University Linz (irene.tubikanec@jku.at)}\\}
	\date{}
	\maketitle	
		
	\thispagestyle{empty}
	
	\vspace{-0.3cm}
	
	\section*{Abstract}
	
	\vspace{-0.2cm}
	
	In this article, we propose an adapted sequential Monte Carlo approximate Bayesian computation (SMC-ABC) algorithm for network inference in coupled stochastic differential equations (SDEs) used for multivariate time series modeling. Our approach is motivated by neuroscience, specifically the challenge of estimating brain connectivity before and during epileptic seizures. To this end, we make four key contributions. First, we introduce a $6N$-dimensional SDE to model the activity of $N$ coupled neuronal populations, extending the (single-population) stochastic Jansen and Rit neural mass model used to describe human electroencephalography (EEG) rhythms, particularly epileptic activity. Second, we construct a reliable and efficient numerical splitting scheme for the model simulation. Third, we apply the proposed adapted SMC-ABC algorithm to the neural mass model and validate it on different types of simulated data.  
	Compared to standard SMC-ABC, our approach significantly reduces computational cost by requiring fewer model simulations to reach the desired posterior region, thanks to the inclusion of binary parameters describing the presence or absence of coupling directions. Finally, we apply our method to real multi-channel EEG data, uncovering potential similarities in patients' brain activities across different epileptic seizures, as well as differences between pre-seizure and seizure periods. 
	
	\vspace{-0.15cm}
	
	\subsubsection*{Keywords} 
	
	\vspace{-0.15cm}
	
	Simulation-based inference, 
	Network inference,
	Approximate Bayesian computation,
	Numerical splitting methods,
	Jansen and Rit neural mass model,
	EEG data
	
	\vspace{-0.15cm}
	
	\subsubsection*{AMS subject classifications} 
	
	\vspace{-0.15cm}
	
	60H10, 60H35, 65C30
	
	\vspace{-0.15cm}
	
	\subsubsection*{Authorship contribution statement} 
	
	\vspace{-0.15cm}
	
	All authors designed the study. M.T. and I.T. wrote the code and M.T. created the R-package. I.T. derived the model and developed the nSMC-ABC algorithm, with support from S.D. and M.T. Moreover, I.T. constructed the numerical splitting method, performed the analysis, produced all figures and wrote the first draft. All authors discussed the results and implications, commented on the manuscript at all stages, edited the text and approved the final version. 
	
	\vspace{-0.15cm}
	
	\subsubsection*{Acknowledgements} 
	
	\vspace{-0.15cm}
	
	A part of this paper was written while I.T. was member of the Department of Statistics, University of Klagenfurt, 9020 Klagenfurt, Austria. The authors are grateful for the high-performance computing infrastructure and the generous mobility funding provided  by that university. S.D. was supported by the Novo Nordisk Foundation NNF20OC0062958.
	
	\vspace{0.2cm}

	
	\section{Introduction}

	Estimating connectivity in a network of units is important in a wide variety of applications, ranging from biology over climate to finance, physics, sociology, and other fields. This is a statistically challenging task, as these interacting units typically follow some complex underlying stochastic dynamics, which may only be partially observed. Moreover, the detection of directed connections and the distinction from spurious ones is particularly difficult. This article is motivated by neuroscience, where the study of neural activity and the underlying connectivity between brain regions is essential for understanding brain function and its implications in various neurological conditions. In particular, we are interested in inferring the connectivity structure of brain regions before and during epileptic seizures, a critical area of study to understand and manage epilepsy. 
	
	Electroencephalography (EEG) is a widely used technique to measure and analyze brain activity, providing insights into the complex dynamics of the brain. Electrodes are placed on the scalp, and the electrical activity is measured at different locations, providing a multidimensional time series. An EEG is used, for example, to find changes in brain activity that might aid in diagnosing epilepsy and other brain conditions. Epilepsy is a neurological disorder that causes recurrent epileptic seizures, characterized by abnormal, excessive, and synchronized electrical discharge in brain neurons, which can be detected by EEG recordings.
	
	Many statistical techniques exist to analyze EEG recordings, mainly nonparametric methods to determine activation areas and other statistical quantities of interest \cite{EEGbook2016}. Inferring the functional network connectivity between EEG channels is of special interest, in particular, to detect differences between normal and seizure behavior. Also in that case, non-parametric inference methods are typically applied, mainly providing correlations that cannot distinguish the  
	coupling directions between brain regions. Here, we aim at estimating the directed connections between $N$ coupled neural populations whose activity corresponds to the simultaneous measurements from $N$ electrodes in the EEG recordings, one per population. Previous works have used cointegration between phases \cite{Ostergaard2017}, cointegration directly in EEG time series \cite{Levakova2022} and multidimensional Ornstein-Uhlenbeck processes \cite{Ruse2019} to infer the directed functional network of the EEG channels. However, they all rely on linear models, whereas more physically based nonlinear models are yet to be explored for statistical inference.
	
	In this article, each neural population is modeled with a stochastic version of the Jansen and Rit neural mass model (JR-NMM) originally proposed in \cite{Jansen1995}. This model has been shown to be useful for reproducing human EEG rhythms, including those associated with epileptic activity. The original JR-NMM is a $6$-dimensional system of ordinary differential equations (ODEs) and describes the average activity of one neural population, corresponding to the measured activity of one electrode in EEG recordings. The model includes a term representing noisy extrinsic input from the neighborhood or more distant regions. As this term can be interpreted as a stochastic process, the solution of the dynamical system is a stochastic process as well, inheriting the analytical properties of this stochastic input function. 
	
	The model was therefore reformulated as a stochastic differential equation (SDE) in~\cite{Ableidinger2017}, and proved to be geometrically ergodic, which guarantees that the distribution of the solution converges to a unique limit distribution, exponentially fast and for any initial value. This has two important statistical implications. First, the choice of the (unknown) initial value is negligible, since its impact on the distribution decreases exponentially fast. Second, quantities related to the distribution can be estimated from a (sufficiently long) single path, avoiding the need for many repeated paths. While the original JR-NMM has been extended to model $N$ coupled neural populations in \cite{Wendling2000}, no such extension has been proposed for the SDE version of this model. We fill in this gap by proposing a $6N$-dimensional SDE model accounting for directed connections between $N$ neural populations (\textit{first contribution}). In contrast to \cite{Wendling2000}, our model contains  $\{0,1\}$-valued coupling direction parameters, describing the underlying network structure. These binary indicators turn out to be crucial for robust and efficient inference of the functional network from EEG data.
	
	As neither the underlying transition density nor exact simulation schemes are available for the $6N$-dimensional SDE model, a suitable numerical approximation is required. The commonly applied Euler-Maruyama discretization is not suitable for this SDE, as it has been shown to fail in preserving crucial structural properties for the single population model, such as the dynamics of the modeled neural oscillations \cite{Ableidinger2017,Buckwar2019}. The interested reader is also referred to \cite{BuckwarTubikanec2022,Chevallier2020,Cohen2012,Kelly2017,Strommen2004,Tubikanec2022}, where similar issues of the Euler-Maruyama method applied to oscillatory SDE models are reported. 
	
	When embedded into a statistical inference procedure, this standard numerical scheme may therefore yield wrong estimation results, make the inference algorithm computationally infeasible or lead to ill-conditioned estimation methods \cite{Buckwar2019,Ditlevsen2019,pilipovic2022parameter,Pokern2009}. To obtain a reliable and efficient numerical simulation method, we further develop the structure-preserving splitting procedure proposed in \cite{Ableidinger2017} (see also \cite{Alamo2017,Skeel1999} for similar methods) to our $N$-population SDE model (\textit{second contribution}). In contrast to the Euler-Maruyama scheme, which is based on truncating a stochastic Taylor series, 
	the idea behind the splitting approach is to divide the unsolvable equation into solvable subequations, and to compose their solutions in a proper way \cite{Blanes2009,Mclachlan2002}. The constructed splitting scheme for the stochastic $N$-population neural mass model is based on a Hamiltonian type re-formulation of the SDE and successfully handles multiple interacting components, allowing to accurately simulate the complex dynamics of coupled neural populations.  
	
	As a next step, we aim to estimate the $\{0,1\}$-valued coupling direction parameters between the $N$ neural populations, as well as relevant real-valued model parameters of the $6N$-dimensional SDE from $N$ simultaneously recorded EEG signals. This is particularly challenging, since this SDE falls into the class of hypoelliptic models \cite{Ditlevsen2019,Iguchi2024,Leon2018,Melnykova2018,pilipovic2024strangsplittingparametricinference,Pokern2009,Samson2025} and is only partially observed via $N$ one-dimensional linear functions of a subset of the $6N$ model components. These issues, combined with the lack of a tractable underlying likelihood, make this problem naturally suitable for likelihood-free inference approaches. 
	
	We focus on the simulation-based Approximate Bayesian Computation (ABC) method \cite{sisson2018handbook}, which has become one of the leading tools for parameter estimation in complex mathematical models in the last decades. 
	The basic ABC algorithm, originally introduced in the context of population genetics \cite{Beaumont2002}, is computationally expensive due to its reliance on parameter candidates  sampled directly from the underlying prior distribution. Since the prior typically concentrates its mass in regions ``far away'' from the posterior, this results in low acceptance rates and high computational costs. To address this, we consider the sequential Monte Carlo (SMC) ABC approach, which represents the state-of-the-art sampler within ABC. Unlike basic ABC, SMC-ABC iteratively constructs targeted proposal samplers, avoiding improbable parameter regions and progressively refining intermediate approximate posterior distributions towards the desired posterior \cite{Beaumont2009,DelMoral2012,Marin2012,Sisson2007}. 
	
	In particular, we propose an adapted SMC-ABC algorithm for network inference, that we call \textbf{nSMC-ABC} (\textit{third contribution}). The nSMC-ABC algorithm efficiently handles high-dimensional parameter spaces by employing two independent proposal samplers: a Gaussian kernel for real-valued parameters and a Bernoulli kernel for $\{0,1\}$-valued network parameters. By leveraging the introduced binary indicators, our method reduces the number of continuous network parameters, significantly lowering the computational costs compared to standard SMC-ABC.
	
	Building upon \cite{Buckwar2019}, where the authors developed an ABC framework for Hamiltonian-type SDEs observed via univariate time series, nSMC-ABC advances the methodology in three key aspects. First, it incorporates the derived numerical splitting scheme for synthetic data generation from the $6N$-dimensional SDE. Second, it constructs effective summary statistics by mapping the $N$-dimensional time series to their estimated $N$ marginal densities and spectral densities, as well as to their estimated 
	cross-correlation functions to account for possible dependencies among the populations. Third and most importantly, it is designed to jointly infer both real-valued model parameters and the $\{0,1\}$-valued network parameters. By estimating the latter, we uncover the directed connectivity structure of  neural populations, enabling the identification of intricate network interactions across brain regions.  
	
	After validating the performance of the proposed statistical approach on different types of simulated data, we apply it to real multi-channel EEG data measured before and during epileptic seizures (\textit{fourth contribution}). We obtain unimodal posteriors of a relatively large number of continuous model parameters and clear network estimates. These results indicate, for example, similarities in patients' brain activities during (or before) epileptic seizures. 
	
	The proposed algorithm, along with the choice of its key ingredients, holds promise beyond the specific application, as it can be applied to other coupled SDEs for stationary time series modeling, provided that a reliable numerical simulation method can be derived. This flexibility broadens the scope of our work, offering a potential avenue for parameter and network estimation in various complex stochastic systems involving coupled units. 
	
	The paper is organized as follows. In Section \ref{sec:2}, we introduce the model. In Section~\ref{sec:4}, we describe the nSMC-ABC algorithm and its required ingredients. In Section \ref{sec:5}, we illustrate its performance on simulated data. In Section \ref{sec:6}, we apply our method to real EEG data with epileptic activity. Conclusions and discussion are reported in Section~\ref{sec:7}. An appendix including the detailed derivation of the splitting method as well as further illustrations of the proposed nSMC-ABC method is attached. 
	Moreover, sample code and a comprehensive \texttt{R}-package are available on GitHub under the links provided in Section \ref{sec:Impl_det}.
	
	\begin{notation}
		We denote by $0_d$ the $d$-dimensional zero vector, by  $\mathbb{O}_d$ the $d\times d$-dimensional zero matrix, by $\mathbb{I}_d$ the $d\times d$-dimensional identity matrix, and by $\textrm{diag}[a_1,\ldots,a_d]$ a $d\times d$-dimensional diagonal matrix with diagonal entries $a_1,\ldots,a_d$. The transpose is denoted by $^\top$ and the Euclidean norm by $\norm{\cdot}$. We sometimes omit the time index of a stochastic process and use, e.g., $(X(t))_{t\in [0,T]}$ and $X$ interchangeably.
	\end{notation} 
	
	
	\section{Stochastic multi-population Jansen and Rit neural mass model}
	\label{sec:2}
	
	In this section, we introduce the stochastic multi-population JR-NMM, whose parameters are then estimated via the nSMC-ABC method proposed in Section \ref{sec:4}. 
	
	In Section \ref{sec:2:1}, we first recall the stochastic JR-NMM of one neural population \cite{Ableidinger2017}. In Section~\ref{sec:2:2}, we then extend this model to a system of multiple coupled neural populations, following the strategies proposed in \cite{Wendling2000}, and introduce binary parameters defining the functional network structure among the neural populations. 
	
	
	\subsection{One neural population}
	\label{sec:2:1}
	
	In the single-population model, a population of neurons (e.g., a cortical column) is described by a system of three interacting sub-groups of neurons. Specifically, the main neurons (pyramidal cells, which form group 1) receive feedback from excitatory (group 2) and inhibitory (group 3) local interneurons (other nonpyramidal cells, stellate or basket cells). 
	
	Each of the three sub-groups is modeled by two blocks. The first block transforms the incoming average firing rate (counting the number of electrical impulses, also known as action potentials or spikes, over time) of a sub-group into an average postsynaptic membrane potential, which is either excitatory or inhibitory. It introduces a second-order ordinary differential operator of the form 
	\begin{equation}\label{eq:first_block}
		\ddot{x}(t)=Ddz(t)-2d\dot{x}(t)-d^2x(t),
	\end{equation} 
	where $z(t)$ and $x(t)$ denote the univariate input and output signal, respectively, $D=A$ (resp. $D=B$) is an excitation (resp. inhibition) parameter and $d=a$ (resp. $d=b$) a time constant. The second block transforms the average incoming membrane potential of a sub-group into an average firing rate. Since the oscillation generation mechanisms are nonlinear, the sigmoid function is used, given by
	$\textrm{sig} \colon \mathbb{R}\to [0,\nu_{\max}]$, $\nu_{\max}>0$, with
	\begin{equation*}
		\textrm{sig}(x):=\frac{\nu_{\max}}{1+e^{\gamma(v_0-x)}},
	\end{equation*}
	where $x$ is a potential, $\nu_{\max}$ describes the maximum firing rate of the neural population, $v_0 \in \mathbb{R}$ is the potential value for which $50\%$ of the maximum firing rate is attained and $\gamma>0$ is proportional to the slope of the sigmoid function at $v_0$. 
	
	Denoting $x_i$, $i \in \{1,2,3\}$, the average postsynaptic membrane potentials (output signals) of the main cells, excitatory and inhibitory interneurons, respectively, yields the following three coupled second-order ODEs
	\begin{align}\label{eq:JR_original}
		\begin{split}
			\ddot{x}_1(t)&= Aa \ \textrm{sig}(x_2(t)-x_3(t))  -2a\dot{x}_1(t)-a^2x_1(t)\\
			\ddot{x}_2(t)&=Aa \left[ p(t)+ C_2  \textrm{sig}(C_1 x_1(t)) \right] -2a\dot{x}_2(t)-a^2x_2(t)\\
			\ddot{x}_3(t)&= Bb C_4  \textrm{sig}(C_3 x_1(t))  -2b\dot{x}_3(t)-b^2x_3(t),
		\end{split}	
	\end{align}
	where $C_1$, $C_2$, $C_3$, $C_4$ are internal connectivity constants, characterizing the interaction between the main cells and the excitatory and inhibitory interneurons, and $p(t)$ is a stochastic input function, modeling excitatory input from neighboring or more distant brain regions.
	
	The meaning and typical values of the model parameters $A$, $B$, $a$, $b$, $C_1$, $C_2$, $C_3$, $C_4$, $\nu_{\max}$, $\gamma$ and $v_0$ are reported in Table \ref{table1} (see \cite{Ableidinger2017,Jansen1995} and the references therein). The average incoming membrane potential of the main cells, i.e., the process
	\begin{equation*}
		\vspace{-0.2cm}
		y(t):=x_2(t)-x_3(t),
		\vspace{-0.2cm}
	\end{equation*}
	describes an EEG signal. 
	
	\begin{table}
		{\footnotesize  
			\caption{Typical parameter values for the Jansen and Rit Neural Mass Model from the literature.}
			\vspace{-0.2cm}
			\label{table1}
			\begin{center}
				\scalebox{1.0}{
					\begin{tabular}{l|l|l}
						\hline 
						Parameter & Meaning & Typical value  \\  
						\hline
						$A$ & Average excitatory synaptic gain & $3.25$ mV  \\					
						$B$ & Average inhibitory synaptic gain & $22$ mV  \\					
						$a$ & Membrane time constant of excitatory postsynaptic potential & $100$ s$^{-1}$ \\					
						$b$ & Membrane time constant of inhibitory postsynaptic potential  & $50$ s$^{-1}$ \\
						$C$ & Average number of synapses between the subpopulations & $135$  \\					
						$C_1$, $C_2$ & Avg. no. of synaptic contacts in the excitatory feedback loop & $C$, $0.8 \ C$   \\					
						$C_3$, $C_4$ &  Avg. no. of synaptic contacts in the inhibitory feedback loop & $0.25 \ C$, $0.25 \ C$  \\
						$\nu_{\max}$ & Maximum firing rate (Maximum of the sigmoid function) & $5$ s$^{-1}$ \\  			
						$v_0$ & Value for which $50\%$ of the maximum firing rate is attained & $6$ mV  \\ 
						$\gamma$ & Determines the slope of the sigmoid function at $v_0$ & $0.56$ mV$^{-1}$ \\ 														
						\hline
				\end{tabular}}
		\end{center}}
	\end{table}  
	
	Since the input function $p(t)$ of the original model \eqref{eq:JR_original} (in \cite{Jansen1995}) essentially is a stochastic process, the model has been re-formulated as an SDE in \cite{Ableidinger2017}, enabling a rigorous mathematical treatment via stochastic (numerical) analysis. In particular, let $(\Omega, \mathcal{F},\mathbb{P})$ be a complete probability space with a complete and right-continuous filtration $(\mathcal{F}(t))_{t\in [0,T]}$, $T>0$. Introducing three further variables, the single-population JR-NMM can be formulated as the following $6$-dimensional~SDE 
	\begin{align}\label{JRNMM_1Pop}
		\begin{split}
			dX_1(t)&=X_4(t)dt\\
			dX_2(t)&=X_5(t)dt\\
			dX_3(t)&=X_6(t)dt\\
			dX_4(t)&=\left[Aa \bigl( \textrm{sig}\left(X_2(t)-X_3(t)\right) \bigr) -2aX_4(t)-a^2X_1(t)\right]  dt+ \bar{\epsilon} dW_4(t)\\
			dX_5(t)&=\left[Aa \bigl(\mu+C_2\textrm{sig}\bigl(C_1 X_1(t)\bigr)\bigr)-2aX_5(t)-a^2X_2(t)\right]dt+\sigma dW_5(t)\\
			dX_6(t)&= \left[BbC_4 \textrm{sig}\left(C_3 X_1(t)\right)-2bX_6(t)-b^2X_3(t)\right]dt+\tilde{\epsilon} dW_6(t),
		\end{split}	
	\end{align}
	for $t\in[0,T]$, with independent Wiener processes $W_i=(W_i(t))_{t\in[0,T]}$, $i=4,5,6$, defined on $(\Omega, \mathcal{F},\mathbb{P})$ and adapted to $(\mathcal{F}(t))_{t\in [0,T]}$, and $\mathcal{F}(0)$-measurable initial value $X_0=(X_1(0),\ldots,X_6(0))^\top$, which is independent of $W=(W_4,W_5,W_6)^\top$ and satisfies $\mathbb{E}\bigl[\norm{X_0}^2\bigr]<\infty$. The parameters $\mu$ and $\sigma$ scale the deterministic and stochastic input, respectively, coming from neighboring or more distant regions in the brain. Together with the Wiener process $W_5$, they thus replace the stochastic input function $p(t)$ of the original model. While usually $\sigma \gg 1$, weak noise acts on the components $X_4$ and $X_6$ with noise intensities $\bar{\epsilon}, \tilde{\epsilon} \ll \sigma$. Throughout, we set $\epsilon=\bar{\epsilon}=\tilde{\epsilon}=1 \ s^{-1}$, since their role is marginal, see \cite{Ableidinger2017} for further details. 
	
	The stochastic single-population JR-NMM \eqref{JRNMM_1Pop} is an additive noise SDE with globally Lipschitz drift coefficient, and thus has a pathwise unique solution $X(t)=(X_1(t),\ldots,X_6(t))^\top$, $t \in [0,T]$,
	which is adapted to $(\mathcal{F}(t))_{t\in [0,T]}$ \cite{Arnold1974,Mao2011}. Moreover, system \eqref{JRNMM_1Pop} is hypoelliptic and geometrically ergodic \cite{Ableidinger2017}. The $6$-dimensional solution $X=(X(t))_{t\in[0,T]}$ is only partially observed through the difference of two of its coordinates
	\begin{equation}
		\label{eq:Y}
		Y(t):=X_2(t)-X_3(t), \quad t \in [0,T].
	\end{equation}
	The discretely observed one-dimensional process $Y=(Y(t))_{t\in[0,T]}$ describes an EEG signal, recorded over a time interval of $T$ seconds. 
	
\begin{figure}
		\centering
		{\includegraphics[width=1.0\textwidth]{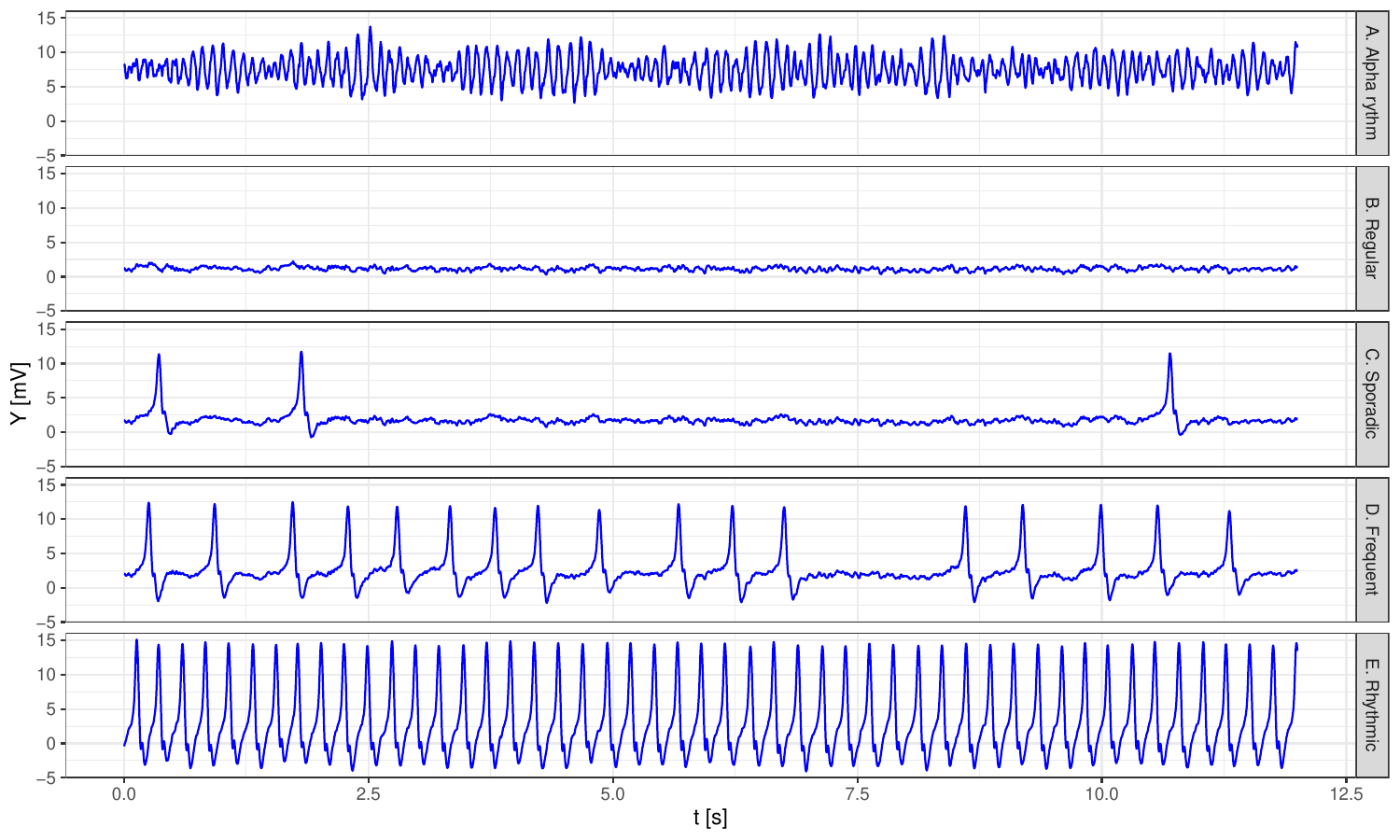}} 
		\caption{{\bf Traces of one neural population with different types of activity.} Simulated paths of the process $Y$ \eqref{eq:Y} of model \eqref{JRNMM_1Pop}. Panel A: Trace showing $\alpha$-rhythmic activity using $C=134.263$, $\mu=202.547$, $\sigma=1859.211$ (values taken from \cite{Buckwar2019}), and $A=3.25$. Panels B-E: Traces describing activity occurring during epileptic seizures using $C=135$, $\mu=90$, $\sigma=500$ and different values of $A$. Panel B: Regular EEG, $A=3.25$. Panel C: Sporadic spikes, $A=3.5$. Panel D: Frequently occurring spikes, $A=3.6$. Panel E: Rhythmic discharge of spikes, $A=4.3$. All remaining parameter values correspond to those listed in Table \ref{table1}. The signals resemble experimental stereo EEG recordings (cf. Figure~3 in \cite{Wendling2000}).
		}
		\label{fig:traces}
	\end{figure}

	An illustration of a simulated trace with parameters chosen to produce $\alpha$-waves (neural oscillations in the 8--12 Hz frequency band) is provided in 
	Figure \ref{fig:traces}A. The model can also produce more complex behavior, such as brain signals occurring before and during epileptic seizures, by increasing the excitation-inhibition-ratio $A/B$ \cite{Wendling2000}. In 
	Figure \ref{fig:traces}B, regular EEG activity is produced by using  $\mu=90$,  $\sigma=500$, $A=3.25$ and the typical values reported in Table~\ref{table1}. Increasing $A$ causes the model to generate sporadic and frequently occurring spikes ($A = 3.5$ and $A=3.6$) and rhythmic discharge of spikes ($A=4.3$), see Figure  \ref{fig:traces}C-\ref{fig:traces}E.
	
	\subsection{Multiple coupled neural populations}
	\label{sec:2:2}
	
	In the multi-population model, initially discussed in \cite{Jansen1995} and more thoroughly investigated in \cite{Wendling2000} in the original setting, $N>1$ neural populations are coupled. A schematic illustration of this model can be found in Figure 1 of \cite{Wendling2000}. Notably, each population receives not only general excitatory external input from its surroundings (modeled by the parameters $\mu$ and $\sigma$, along with the Wiener process $W_5$), but also afferent excitatory input derived from the average firing rate of main cells in other populations. Specifically, the $k$-th population, for $k\in \{1,\ldots,N\}$, receives input from $X_1^j(t)$, for $j\neq k$, via the fifth equation according to the first block (cf. \eqref{eq:first_block}). This is biologically motivated by the fact that the main pyramidal cells (whose output is modeled by $X_1^j(t)$) are excitatory neurons with axons that project to other brain~regions. 
	
	Incorporating this connectivity structure into the model yields a $6N$-dimensional SDE, 
	where the $k$-th population $X^k(t)=(X_1^k(t),\ldots, X_6^k(t))^\top$, for $t\in [0,T]$, satisfies system \eqref{JRNMM_1Pop} with suitable index $k$, except for the fifth equation, which is given~by
	{\small{
	\begin{equation}\label{eq:coupling}
			 dX^k_5(t)=\Bigl[A_ka_k \Bigl(\mu_k+C_{2,k}\textrm{sig}\left(C_{1,k} X^k_1(t)\right) +\hspace{-3mm} \sum\limits_{j=1,j\neq k}^{N} \rho_{jk}K_{jk} X_1^j(t) \Bigr)-2a_kX^k_5(t)-a_k^2X^k_2(t)\Bigr]dt
				+\sigma_{k} dW^k_5(t). 
	\end{equation}}}\noindent
	The new coupling term in \eqref{eq:coupling} contains both binary parameters $\rho_{jk}\in \{ 0,1 \}$, which determine whether there is ($\rho_{jk}=1$) or not ($\rho_{jk}=0$) a directed coupling from the $j$-th to the $k$-th population, and continuous parameters $K_{jk}>0$, which model the eventual coupling strength from population $j$ to $k$. 
	
	
	These additional binary coupling direction parameters $\rho_{jk}\in \{ 0,1 \}$ in \eqref{eq:coupling}, not present in \cite{Wendling2000}, are essential for network inference via ABC, as they allow for a significant reduction of the number of continuous parameters, and, thus, of the computational cost of the proposed nSMC-ABC algorithm (see Section~\ref{sec:4} for a detailed description of the algorithm and Appendix \ref{app:binaryIndicators} for a detailed  illustration of these aspects). In particular, to reduce the number of continuous parameters in the model, we assume that the coupling strength between two populations decreases with increasing population distance, where the distance is defined by the difference between their subindices. In particular, for $j,k \in \{1,\ldots,N\}$ with $j\neq k$, define 
	\begin{equation}\label{eq:coupling_structure}
		K_{jk}:=c^{|j-k|-1}L,
	\end{equation}
	where $L>0$ is a coupling strength parameter and the parameter $0 \ll c < 1$ describes how fast the network coupling strength decreases with increasing distance between populations. 
	
	\begin{remark}
		Coupling terms with parameters $\rho_{jk}\in \{0,1\}$ and $K_{jk}>0$ as in \eqref{eq:coupling} and \eqref{eq:coupling_structure} can also be used to define functional networks for other SDE models.
	\end{remark}
	
	
	The parameters of the stochastic multi-populatuion JR-NMM thus consist of continuous model parameters (cf. Table \ref{table1}), two further continuous coupling strength parameters $L>0$ and $\text{$0 \ll c < 1$}$ (cf. \eqref{eq:coupling} and \eqref{eq:coupling_structure}) and binary coupling direction (network) parameters $\rho_{jk}$ $j,k=1,\ldots,N$, $j \neq k$ (cf. \eqref{eq:coupling}).
	Moreover, the observed component of the model is given by the $N$-dimensional process 
	\begin{equation}\label{eqn:output}
		Y(t):=(Y^1(t),\ldots,Y^N(t))^\top=(X^1_2(t)-X^1_3(t),\ldots,X^N_2(t)-X^N_3(t))^\top, \quad t\in[0,T],	
	\end{equation}
	which describes $N$ EEG signals, simultaneously recorded during $T$ seconds. 
	
\begin{figure}
		\begin{centering}
			\includegraphics[width=1.0\textwidth]{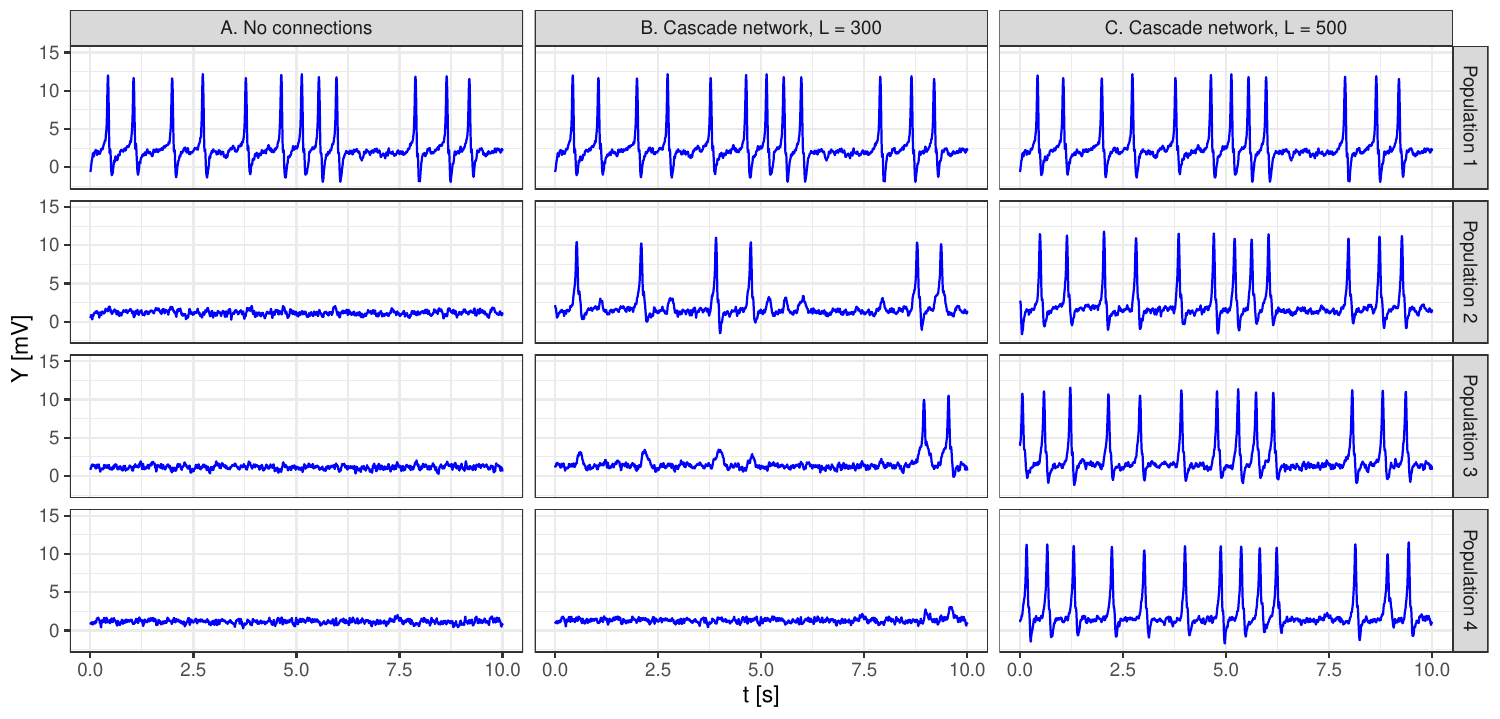}	
			\caption{{\bf Cascade network of four populations with one active population.} Simulated paths of $Y$~\eqref{eqn:output} for $N=4$ neural populations. The typical values of Table~\ref{table1}, $\mu=90$ and $\sigma=500$ are used, except for Population $1$ where $A_1=3.6$ to activate it. Panel~A: $\rho_{jk} =0$, $j,k=1,\ldots,4$, $j\neq k$. Panels B-C: $\rho_{12}=\rho_{23}=\rho_{34}=1$. The coupling strength parameters $K_{12}$, $K_{23}$ and $K_{34}$ equal $L=300$ and $L=500$ in Panel B and Panel C, respectively.}
			\label{fig:paths_coupling}
		\end{centering}
	\end{figure}

	
	Similar to Section \ref{sec:2:1}, we present a short simulation study where we generate paths of $Y$ \eqref{eqn:output}, focusing on the newly introduced coupling term. According to \cite{Wendling2000}, not only the excitation-inhibition-ratio $A/B$ is relevant for epileptic behavior, but also the coupling strengths and directions between neural groups play a crucial role. This is illustrated in Figure \ref{fig:paths_coupling}, where simulated activity of $N=4$ neural populations under different coupling regimes is shown. In the left panels, no coupling occurs, i.e., all $\rho_{jk}$-parameters are set to zero. In the middle and right panels, there is a unidirectional (cascade) coupling structure (illustrated in Figure \ref{fig:network_structures_N4}a of Section~\ref{sec:5}), i.e, $\rho_{12}=\rho_{23}=\rho_{34}=1$, for different coupling strengths $L=K_{12}=K_{23}=K_{34}$. The activity of a passive site (no epileptic spikes occur without input from other populations) strongly depends on that of an active site (epileptic spikes occur without input from other populations). In Population~$1$, we set $A_1=3.6$ to obtain spiking activity. In the remaining populations, the typical values of Table~\ref{table1}, $\mu=90$ and $\sigma=500$ are used. When there is no coupling, no activation of Populations~$\text{2--4}$ occurs (left panels). Introducing coupling and setting  $L=K_{12}=K_{23}=K_{34}=300$ (central panels) leads to a dependence of Populations 2--4 on Population $1$. When the coupling is strong enough ($L=K_{12}=K_{23}=K_{34}=500$, right panels), rhythmic synchronization occurs. A similar behavior for two populations has also been observed for the original JR-NMM (cf. Figure~6 in \cite{Wendling2000}).
	
	\section{Adapted SMC-ABC algorithm for continuous and binary parameters}
	\label{sec:4}
	
	The stochastic multi-population JR-NMM detailed in the previous section is representative for the class of partially observed models with unknown likelihood function, containing both continuous (real-valued) parameters and binary ($\{0,1\}$-valued) network parameters. In this section, we  propose an adapted SMC-ABC algorithm for this class of models (the nSMC-ABC algorithm), where the Gaussian proposal sampler within the standard SMC-ABC algorithm is extended by a Bernoulli-type kernel to account for both continuous and binary parameters.  
	
	In Section \ref{sec:ABCschemes}, we first review the standard acceptance-rejection, reference-table acceptance-rejection, and SMC-ABC schemes. In Section \ref{sec:nSMC-ABC}, we introduce the proposed nSMC-ABC algorithm. In Section \ref{sec:SMC_ABC_network}, we adapt it to the stochastic multi-population JR-NMM, detailing all the required model-specific algorithmic ingredients (numerical scheme for synthetic data generation, suitable summary statistics and distance measures) as well as relevant implementation details.
	
	
	\subsection{Standard ABC schemes}
	\label{sec:ABCschemes}
	
	Let $\theta$ denote the parameter vector of the model of interest, to be inferred from the observed data $y$. Denoting by 
	$\pi(\theta)$ the prior and by 
	$\mathcal{L}(y|\theta)$ the likelihood function, the posterior distribution $\pi(\theta|y)$ satisfies 
	\begin{equation*}
		\pi(\theta|y) \propto \pi(\theta)\mathcal{L}(y|\theta).
	\end{equation*}
	In general, the likelihood function, and thus the true posterior, are not available for complex models, such as the stochastic multi-population JR-NMM. The idea of ABC is to replace the unknown likelihood via a large amount of \lq\lq synthetic\rq\rq\ datasets simulated from the model, obtaining an approximate posterior distribution $\pi_\textrm{ABC}(\theta|y)$ targeting the true (unavailable) posterior $\pi(\theta|y)$.
	
	Several ABC algorithms have been proposed, see \cite{sisson2018handbook} for an overview. The simplest is \textit{acceptance-rejection ABC} \cite{Beaumont2002,Marin2012,sisson2018handbook}, consisting of three steps: (a) Sample $\theta^{'}$ from the prior $\pi(\theta)$; (b) Conditioned on $\theta^{'}$,  simulate a synthetic dataset $\tilde{y}_{\theta^{'}}$ from the model; (c) Keep the sampled value $\theta^{'}$ if 
	the distance $D(\cdot,\cdot)$ between a vector of summary statistics $s(\cdot)$ of the observed and simulated data is smaller than a threshold $\delta>0$, i.e., $
	D(s(y),s(\tilde y_{\theta'}))<\delta$. Steps (a)-(c) are then repeated until $M$ draws are accepted, which typically happens after $n\gg M$ drawings. This leads to the approximate acceptance-rejection ABC posterior 
	\[\pi(\theta|y)\approx \pi_{\textrm{ABC}}^\delta(\theta|s(y))\propto \int \mathbbm{1}_{\{D(s(y),s(\tilde y_{\theta'}))<\delta\}}\pi(\theta)\mathcal{L}(s(\tilde y_{\theta'})|\theta)ds.
	\]
	Instead of keeping only the samples whose distance is smaller than some apriori fixed threshold~$\delta$, the \textit{reference table acceptance-rejection ABC scheme} \cite{Buckwar2019,Cornuet2008} first produces a \textit{reference table} $\{\theta_l,D_l\}$, $l=1\ldots,n$, and then selects the threshold level $\delta$ as the $q$-th percentile of the distances $D_l$. This procedure has the practical advantage of fixing the number of model simulations $n$ in advance, and it is often used to perform a pilot ABC study for selecting an initial threshold $\delta_1$ required in sequential ABC algorithms.
	
	Acceptance-rejection ABC and its variant are computationally inefficient by construction, as the proposals $\theta'$ are sampled from the prior distribution throughout, with a high computational waste. To tackle this, here we consider \textit{SMC-ABC} \cite{Beaumont2009,DelMoral2012,Sisson2007}, a sequential ABC algorithm. It runs through multiple iterations $r=1,\ldots,r_\textrm{last}$ using so-called proposal samplers (also known as perturbation kernels or important samplers), which are based on the kept sampled values (called \textit{particles}) at the previous iteration. This yields intermediate approximate posterior distributions, which move closer and closer to the desired posterior during consecutive iterations, accelerating the estimation procedure. In particular, SMC-ABC works as follows.
	
	At iteration $r=1$, acceptance-rejection ABC is run, sampling
	from the prior distribution until $M$ 
	particles $\Theta_1=(\theta_1^{(1)},\ldots,\theta_1^{(M)})$
	have been accepted, i.e., have yielded a distance smaller than an initial threshold $\delta_1$. Then, the initial particle weights are set to $w_1=(w_1^{(1)},\ldots, w_1^{(M)})=(1/M,\ldots, 1/M)$. 
	
	At iteration $r>1$, a particle $\theta$ is initially sampled from the set of kept candidates $\Theta_{r-1}$ of the previous iteration with the corresponding weights $w_{r-1}$ and then perturbed to a value $\theta^* \sim \mathcal{K}_r(\cdot | \theta)$, where $\mathcal{K}_r$ is a suitable perturbation kernel.  For continuous parameters $\theta$, $\mathcal{K}_r$  is commonly assumed to be Gaussian (see, e.g., \cite{DelMoral2012,Filippi2013,Samson2025}), even though other possibilities have been proposed \cite{PicchiniTamborrino2022}. Here, we consider the multivariate Gaussian sampler as proposed in \cite{Filippi2013}, that is a candidate $\theta^*$ is obtained from $\mathcal{N}(\theta,\hat{\Sigma}_{r})$, where $\hat{\Sigma}_{r}$ is twice the weighted empirical covariance matrix obtained from the previous population 
	$\{\Theta_{r-1},w_{r-1} \}$ (cf. Appendix \ref{app:perturbationKernel} for an alternative Gaussian proposal sampler). Synthetic data $\tilde{y}_{\theta^*}$ are then simulated conditioned on the perturbed particle  $\theta^*$, which is accepted if $d(s(y),s(\tilde{y}_{\theta^*}))<\delta_r$, with $\delta_r<\delta_{r-1}$. This is repeated until $M$ particles $\Theta_r=(\theta_r^{(1)},\ldots,\theta_r^{(M)})$ have been accepted. In particular, after the $h\text{-th}$ particle $\theta_r^{(h)}$, for $h\in \{ 1,\ldots,M \}$, has been accepted, the corresponding important weight $w_r^{(h)}$ is updated as 
	\[
	\tilde{w}_{r}^{(h)}=\pi\left(\theta_{r}^{(h)}\right) / \sum\limits_{l=1}^{{M}} w_{r-1}^{(l)} \mathcal{K}_r\left( \theta_{r}^{(h)} \Big| \theta_{r-1}^{(l)} \right),
	\]
	where
	{\small{\begin{equation}\label{kernel_c_mvn}
				\mathcal{K}_r\left( \theta_{r}^{(h)} \Big| \theta_{r-1}^{(l)} \right)
				=(2\pi)^{-d/2}\left( \det \hat{\Sigma}_{r} \right)^{-1/2} \exp\left( -\frac{1}{2} \left[  \theta_{r}^{(h)} - \theta_{r-1}^{(l)}  \right]^\top \hat{\Sigma}_{r}^{-1} \left[  \theta_{r}^{(h)} - \theta_{r-1}^{(l)}  \right] \right).
	\end{equation}}}\noindent
	After all important weights have been updated, the $h$-th update $\tilde{w}_{r}^{(h)}$ is normalized~via
	\[
	{w}_{r}^{(h)} = \tilde{w}_{r}^{(h)}/ \sum\limits_{l=1}^{M} \tilde{w}_r^{(l)}, \quad h=1,\ldots,M.
	\]
	This procedure is then repeated over several iterations (with decreasing threshold $\delta_r<\delta_{r-1}$ chosen as a certain percentile of the distances computed at the previous iteration $r-1$) until a suitable stopping criterion is reached (e.g., the acceptance rate of particles drops below a prefixed level or a certain number of model simulations, denoted by $n_{\textrm{sim}}$, has been reached). The final SMC-ABC posterior is obtained by sampling from the particles kept at the last iteration $r_\textrm{last}$ with probabilities given by the corresponding normalized weights. A detailed formulation of the standard SMC-ABC can be found, e.g., in Algorithm~2 of \cite{PicchiniTamborrino2022}.
	
	
	\subsection{Adapted SMC-ABC algorithm for network inference: nSMC-ABC}\label{AdjustedSMCABC}
	\label{sec:nSMC-ABC}
	
	Here, we adapt the standard SMC-ABC algorithm with Gaussian proposal samplers for continuous model parameters to the specific case of additional binary network parameters. The resulting nSMC-ABC method is reported in  Algorithm~\ref{alg:SMC_SBP_ABC}. 
	
	Specifically, in the nSMC-ABC Algorithm \ref{alg:SMC_SBP_ABC}, we split the parameter vector $\theta$ of the model of interest as
	\begin{equation*}
		\theta:=(\theta_c,\theta_b),
	\end{equation*}
	where $\theta_c$ contains continuous (real-valued) model parameters, while $\theta_b$ consists of the binary ($\{0,1\}$-valued) network parameters. We denote by $c_n$ and $b_n$ the dimensions of these two parameter vectors, respectively, and their entries by $\theta_c^{u}$, ${u}=1,\ldots,c_n$, and $\theta_b^{u}$, ${u}=1,\ldots,b_n$. We write $\Theta_{c,r}$ and $\Theta_{b,r}$ to denote the $M$ kept continuous and binary particles, respectively, at iteration $r$, where each particle $\theta_r^{(h)}$ is represented as
	\begin{equation*}
		\theta_r^{(h)}=\bigl(\theta_{c,r}^{(h)},\theta_{b,r}^{(h)}\bigr), \quad h=1,\ldots,M.
	\end{equation*}
	
	At iteration $r=1$, we run the standard acceptance-rejection ABC, sampling parameter candidates $\theta_c$ and $\theta_b$ from the priors $\pi^c$ and $\pi^b$, respectively (lines $2-10$ of Algorithm \ref{alg:SMC_SBP_ABC}). 
	At iteration $r>1$, as done in standard SMC-ABC, we sample a particle $\theta_{c,r}^{(h)}=\theta_c$ for the continuous parameters from the weighted set $\{\Theta_{c,r-1},{w}_{r-1}\}$ (line~17 of Algorithm \ref{alg:SMC_SBP_ABC}). It is then perturbed to $\theta_c^{*}$ with a Gaussian proposal kernel $\mathcal{K}^c_r$ (defined as in~\eqref{kernel_c_mvn}) and proposed in \cite{Filippi2013} (line 18 of Algorithm~\ref{alg:SMC_SBP_ABC}). In particular, $\theta_c^{*}$ is a realization of $\mathcal{N}(\theta_c,\hat{\Sigma}_{c,r})$, where $\hat{\Sigma}_{c,r}$ is twice the weighted covariance matrix obtained from the previous population $\{\Theta_{c,r-1},{w}_{r-1}\}$ of continuous parameters. Moreover, differently from the classical SMC-ABC, each entry $\theta_b^{u}$ of a particle $\theta_{b,r}^{(h)}=\theta_b$ for a binary parameter is drawn from a Bernoulli distribution with ``success'' probability given by the sample mean obtained from the respective particles of previous population $\Theta_{b,r-1}$ (line 19 of Algorithm~\ref{alg:SMC_SBP_ABC}). Such sampled value $\theta_b^{u}$, for $u=1,\ldots,b_n$, is then perturbed to $\theta_b^{*,u}$ with a Bernoulli-type kernel (line 20 of Algorithm \ref{alg:SMC_SBP_ABC}). In particular, $\theta_b^{u}$ is either kept, i.e., $\theta_b^{*,{u}}=\theta_b^{u}$ (with probability $q_{\textrm{stay}}$) or perturbed to $\theta_b^{*,{u}}=1-\theta_{b}^{u}$ (with probability $1-q_{\textrm{stay}}$). An explicit expression of such kernel $\mathcal{K}_r^b$ (in the style of \eqref{kernel_c_mvn}) is given by 
	\begin{equation}\label{kernel_d2}
		\mathcal{K}_r^b\left( \theta_{b,r}^{(h)} \Big| \theta_{b,r-1}^{(l)} \right) = \prod\limits_{{u}=1}^{b_n}  \mathcal{K}_r^{b,{u}}\left( \theta_{b,r}^{{u},(h)} \Big| \theta_{b,r-1}^{{u},(l)} \right)= 
		\prod\limits_{{u}=1}^{b_n} \left( p_r^{{u},(l)} \right)^{  \theta_{b,r}^{{u},(h)} } \left(1-p_r^{{u},(l)}\right)^{ 1- \theta_{b,r}^{{u},(h)}  },
	\end{equation}
	where
	\begin{equation*}
		p_r^{{u},(l)} = \begin{cases}
			q_{\textrm{stay}}, & \text{if } \theta_{b,r-1}^{{u},(l)} =1 \\
			1-q_{\textrm{stay}}, & \text{if } \theta_{b,r-1}^{{u},(l)} =0
		\end{cases}.
	\end{equation*}
	Throughout this work, the probability $q_\textrm{stay}$ is fixed across iterations (cf. Section \ref{sec:Impl_det}). Alternatively, such probability could also vary, e.g., it may depend on some statistics of the previous population or the number of iterations. We refer to Appendix \ref{app:perturbationKernel} 
	for an investigation of the hyperparameter $q_{\textrm{stay}}$ and alternative proposal samplers for nSMC-ABC.

		\begin{algorithm}[H]
		\caption{Adapted SMC-ABC for network inference: \textbf{nSMC-ABC} 
			\ \\ \textbf{Input:} Summaries $s(y)$ of the observed data $y$, prior distributions $\pi^c$ and $\pi^b$, perturbation kernels $\mathcal{K}_{r}^c$ and $\mathcal{K}_r^b$, number of kept samples per iteration $M$, initial threshold~$\delta_1$ 
			\ \\ 
			\textbf{Output:} Weighted particles from the nSMC-ABC posterior 
		}\label{alg:SMC_SBP_ABC}
		\begin{algorithmic}[1]
			\State Set $r=1$ and $n_{\textrm{sim}}=0$
			\For{$h=1:M$}
			\Repeat
			\State Sample $\theta_b$ from $\pi^b$ and $\theta_c$ from $\pi^c$, and set $\theta=(\theta_{c},\theta_{b})$ 
			\State Conditioned on $\theta$, simulate a synthetic dataset $\tilde{y}_{\theta}$ from the model,\newline and set $n_{\textrm{sim}}=n_{\textrm{sim}}+1$ 
			\State Compute the summaries $s(\tilde{y}_{\theta})$ 
			\State Calculate the distance  $D=d\bigl(s(y),s(\tilde{y}_{\theta})\bigr)$ 
			\Until{$D<\delta_1$}
			\State Set $\theta_{b,1}^{(h)}=\theta_b$ and $\theta_{c,1}^{(h)}=\theta_c$ 
			\EndFor
			\State Initialize the weights by setting each entry of $w_1=(w_{1}^{(1)},\ldots,w_{1}^{({M})})$ to $1/{M}$  
			\Repeat
			\State Set $r=r+1$
			\State Determine $\delta_r< \delta_{r-1}$
			\For{$h = 1:M$}
			\Repeat 
			\State Sample $\theta_c$ from the weighted set $\{ \Theta_{c,r-1}, {w}_{r-1} \}$ 
			\State Perturb $\theta_c$ to obtain $\theta_c^{*}$ from $\mathcal{K}^c_r(\cdot|\theta_c)$
			\State Sample $\theta_b^k$, $k=1,\ldots,b_n$, from $\textrm{Bernoulli}(\hat{p}_r^k)$, where  
			$\hat{p}_r^k= \frac{1}{M} \sum\limits_{l=1}^{M}  \theta^{k,(l)}_{b,r-1}$
			\State Perturb $\theta_b=(\theta_b^1,\ldots,\theta_b^{b_n})$ to obtain $\theta_b^{*}$ from $\mathcal{K}^b_r(\cdot|\theta_b)$ 
			\State Conditioned on $\theta^*=(\theta_c^*,\theta_b^*)$, simulate a synthetic dataset $\tilde{y}_{\theta^*}$ from the model,\newline and set $n_{\textrm{sim}}=n_{\textrm{sim}}+1$  
			\State Compute the summaries $s(\tilde{y}_{\theta^*})$ 
			\State Calculate the distance  $D=d\bigl(s(y),s(\tilde{y}_{\theta^*})\bigr)$ 
			\Until{$D<\delta_r$}
			\State Set $\theta_{b,r}^{(h)}=\theta_b^*$ and $\theta_{c,r}^{(h)}=\theta_c^*$ 
			\State Set  $\tilde{w}_{r}^{(h)}=\pi^c\left(\theta_{c,r}^{(h)}\right) / \sum\limits_{l=1}^{{M}} w_{r-1}^{(l)} \mathcal{K}_r^c\left( \theta_{c,r}^{(h)} \Big| \theta_{c,r-1}^{(l)} \right)$  
			\EndFor
			\State Normalize the weights $w_{r}^{(h)}=\tilde{w}_{r}^{(h)}/ \sum\limits_{l=1}^{{M}} \tilde{w}_{r}^{(l)} $, for $h=1,\ldots,{M}$ 
			\Until{stopping criterion is reached}
			\State Return the final $ \Theta_{b,r_\textrm{last}}$ and $\{ \Theta_{c,r_\textrm{last}}, w_{r_\textrm{last}}\}$. 
		\end{algorithmic}
	\end{algorithm}

	\begin{remark}
		In the proposed nSMC-ABC method, the components $\theta_{c,r}^{(h)}$ and $\theta_{b,r}^{(h)}$ of the $h$-th 
		particle are sampled and perturbed independently. One may also construct an algorithm where a full particle $\theta_r^{(h)}=(\theta_{c,r}^{(h)},\theta_{b,r}^{(h)})$ is first sampled jointly and then perturbed, preserving thus some dependency among the continuous and binary parameters. In Appendix  \ref{app:perturbationKernel}, we investigate such alternative approach, illustrating that it requires far more model simulations (and thus higher computational cost) to reach the desired posterior parameter regions.    
	\end{remark}
	
	\vspace{-0.3cm}
	\begin{remark}
		Algorithm \ref{alg:SMC_SBP_ABC} is related to ABC for model selection as discussed in \cite{Toni2009}, in the sense that each possible network (for a given combination of the binary parameters) may be interpreted as a model. Their algorithm then samples continuous parameter candidates conditioned on a given model and obtains the posterior distribution of the model based on the number of particles kept under each model. Such approach is suitable when there are few possible models, but not here, where we would obtain a prohibitive number of models ($2^{N(N-1)}$ models, e.g., $4096$ for $N=4$).
	\end{remark}
	
	
	\subsection{Adaptation of nSMC-ABC to the stochastic multi-population JR-NMM}
	\label{sec:SMC_ABC_network}
	
	The accuracy and performance of any ABC algorithm, including the proposed nSMC-ABC, depend on various further model-specific aspects, in particular the numerical method used to simulate the synthetic model data, the data summaries and their distances. In the following, we describe our proposed choices of these key ingredients for the multi-population stochastic JR-NMM, which are based on its properties.
	
	\subsubsection{Algorithm ingredients}
	
	\paragraph*{Choice of simulation method}
	
	The standard Euler-Maruyama method cannot be used for the simulation of the stochastic JR-NMM \cite{Ableidinger2017,Buckwar2019}. Therefore, we construct a numerical splitting method for the simulation of the stochastic $N$-population JR-NMM (i.e., synthetic datasets from Y \eqref{eqn:output}), further developing the method presented in \cite{Ableidinger2017} for single neural populations to the multi-population case. The proposed splitting scheme is summarized in Algorithm~\ref{Algorithm_Splitting}, with a detailed derivation  provided in Appendix \ref{app:splitting}.  
	
	Let $0=t_0<\ldots <t_m=T$ be a partition of the time interval $[0,T]$ with equidistant time steps $\Delta=t_{i+1}-t_i>0$, for $i=0,\ldots,m-1$, $m \in \mathbb{N}$. We aim to construct reliable approximations $\widetilde{X}(t_i)$ of the (unknown) process $X(t_i)$ at discrete time points $t_i=i\Delta$, $i=0,\ldots,m$. To do so, define the $3N\times 3N$-dimensional diagonal matrices 
	\begin{equation*}
		\Gamma:=\text{diag}[a_1,a_1,b_1,\ldots,a_N,a_N,b_N], \quad \Sigma:=\text{diag}[\epsilon_{1},\sigma_{1},\epsilon_1,\ldots,\epsilon_N,\sigma_{N},\epsilon_N],
	\end{equation*}
	and the $6N\times 6N$-dimensional matrices
	{\small{\begin{align}\label{eq:ExpCov}
			\begin{split}
				\textrm{Exp}(\Delta)&:=\begin{pmatrix}
					e^{-\Gamma \Delta} \left( \mathbb{I}_{3N}+\Gamma \Delta \right) & e^{-\Gamma \Delta} \Delta \\
					-\Gamma^2 e^{-\Gamma \Delta} \Delta & e^{-\Gamma \Delta} \left( \mathbb{I}_{3N}-\Gamma \Delta \right)
				\end{pmatrix}=:
				\begin{pmatrix}
					\vartheta(\Delta) & \kappa(\Delta)\\
					\vartheta'(\Delta) & \kappa'(\Delta)
				\end{pmatrix},\\
				\textrm{Cov}(\Delta)&:=
				\begin{pmatrix}
					\frac{1}{4} \Gamma^{-3} \Sigma^2 \left( \mathbb{I}_{3N} + \kappa(\Delta)\vartheta'(\Delta)-\vartheta^2(\Delta) \right) & \frac{1}{2}\Sigma^2 \kappa^2(\Delta) \\
					\frac{1}{2}\Sigma^2 \kappa^2(\Delta) & \frac{1}{4} \Gamma^{-1} \Sigma^2 \left( \mathbb{I}_{3N} + \kappa(\Delta)\vartheta'(\Delta)-\kappa'^2(\Delta) \right)
				\end{pmatrix}.
			\end{split}
	\end{align}}}\noindent 
	Moreover, define the function  
	$G(X)=(G_1({X}),\ldots,G_N({X}))^\top$, where for $k \in \{ 1,\ldots,N \}$, 
	\begin{equation}\label{eq:G}
		G_k({X(t)})= \begin{pmatrix}
			A_ka_k \textrm{sig}\left(X^k_2(t)-X^k_3(t)\right) \\
			A_ka_k\Bigl(\mu_k+C_{2,k}\textrm{sig}\left(C_{1,k} X_1^k(t)\right) + \sum\limits_{j=1,j\neq k}^{N} \rho_{jk}K_{jk} X_1^j(t) \Bigr) \\
			B_kb_kC_{4,k}\textrm{sig}(C_{3,k} X_1^k(t))
		\end{pmatrix}.
	\end{equation}
	A path from the stochastic multi-population JR-NMM (and thus a synthetic dataset from the observed process $Y$~\eqref{eqn:output}) can then be simulated via Algorithm \ref{Algorithm_Splitting} (cf. Appendix  \ref{app:splitting} for the details).
	
	\begin{algorithm}
		\caption{Splitting scheme for the stochastic multi-population JR-NMM
			\ \\ \textbf{Input:} Number of populations $N$, step size $\Delta>0$, time horizon $T>0$, initial value $X_0 \in \mathbb{R}^{6N}$, model parameters $\theta$, functions $\textrm{Exp}$ and $\textrm{Cov}$ \eqref{eq:ExpCov}, function $G$~\eqref{eq:G}. \ \\
			\textbf{Output:} Simulated path $(\widetilde{X}(t_i))_{i=0}^{m}$ with synthetic dataset $\tilde{y}_\theta= (\widetilde{Y}(t_i))_{i=0}^{m}$ from \eqref{eqn:output}.}
		\label{Algorithm_Splitting}
		\begin{algorithmic}[1]
			\State Set $\widetilde{X}(t_0)=X_0$ and $m=T/\Delta$
			\For{$i = 0:({m-1})$}
			\State Generate $\xi_i(\Delta)$ from a $6N$-dimensional normal distribution with $\mathcal{N}\left(0_{6_N},\textrm{Cov}(\Delta)\right)$
			\State Set $X^{\textrm{[2]}}=\widetilde{X}(t_i)+\frac{\Delta}{2} \begin{pmatrix}
				0_{3N} \\
				G(\widetilde{X}(t_i))
			\end{pmatrix}$
			\State Set $X^{\textrm{[1]}}=\textrm{Exp}(\Delta) X^{[2]}+\xi_i(\Delta)$
			\State Set $\widetilde{X}(t_{i+1})=X^{[1]}+\frac{\Delta}{2} \begin{pmatrix}
				0_{3N} \\
				G(X^{[1]})
			\end{pmatrix}$
			\EndFor
			\State Return $\widetilde{X}(t_i)$, for $i=0,\ldots,m$, and use $\tilde{y}_\theta=(\widetilde{Y}(t_i))_{i=0}^{m}$ \eqref{eqn:output} for inference.
		\end{algorithmic}
	\end{algorithm}

	\paragraph*{Choice of summary statistics}
	
	An observed dataset $y$ for the stochastic $N$-population JR-NMM corresponds to a multivariate time series dataset, i.e., it consists of $N$ simultaneously recorded univariate time series $y^k=(y^k(t_i))_{i=0}^{m}$, representing observations of the process $(Y^k(t))_{t \in [0,T]}$ in~\eqref{eqn:output}, for $k=1,\ldots,N$. 
	
	Our proposed summary statistics of $y$ consist of functions that capture both the marginal behavior of each population $Y^k$, $k = 1,\ldots,N$, 
	and the interactions between populations $Y^k$ and $Y^j$, $j,k = 1,\ldots,N$, $j\neq k$.
	
	For the marginal behavior of each population, we use the summary statistics proposed in \cite{Buckwar2019} for univariate time series datasets. In particular, taking advantage of the underlying geometric ergodicity of the stochastic JR-NMM, we consider the marginal invariant densities, denoted by $f_k(y^k)$, $k = 1,\ldots,N$, and the marginal spectral densities, given by the Fourier transform
	\begin{equation*}
		S_{k}(\nu)=\mathcal{F}\{ R_k \}(\nu)=\int\limits_{-\infty}^{\infty} R_k(\tau)e^{-i 2\pi \nu \tau} \ d\tau, \quad k=1,\ldots,N,
	\end{equation*}
	where $R_k(\tau)=\mathbb{E}[Y^k(t)Y^k(t+\tau)]$
	is the auto-correlation function of $Y^k$, and $\tau$ and $\nu$ denote the time-lag and frequency, respectively.
	
	To detect possible interactions among $Y^k$ and $Y^j$, $j\neq k$, we additionally consider the cross-correlation functions
	\vspace{-0.3cm}
	\begin{align*}
		R_{jk}(\tau)&=\mathbb{E}[Y^j(t)Y^k(t+\tau)], \quad j, k =1,\ldots,N, \ j\neq k,
	\end{align*}
	with the property that $R_{jk}(\tau)=R_{kj}(-\tau)$. Since they are not symmetric in $j,k$ these summaries enable the detection of directed connections. 
	
	\begin{remark}
		The cross-spectral densities $S_{jk}(\nu)=\mathcal{F}\{ R_{jk} \}(\nu)=\int\limits_{-\infty}^{\infty} R_{jk}(\tau)e^{-i 2\pi \nu \tau} \ d\tau$, $j, k =1,\ldots,N$, $j< k,$ could be other suitable summary statistics. However, these are symmetric (i.e., $S_{jk}(\nu)=S_{kj}(\nu)$), and thus cannot determine the direction of a connection. Indeed, our experiments show that adding those functions to the set of summaries does not yield an additional~benefit.
	\end{remark}
	
	The summary functions $f_k,S_k$ and $R_{jk}$ are estimated from a dataset $y$ via standard estimation procedures (cf.  Section \ref{sec:Impl_det}). Denoting these estimates by $\hat{f}_{k}, \hat{S}_{k}$ and $\hat{R}_{jk}$, the set of summaries $s(\cdot)$ of a dataset $y$ is defined as
	\vspace{-0.3cm}
	\begin{equation}\label{eq:ABC_summaries}
		s(y):=\left\{ \hat{f}_{k}, \hat{S}_{k}, \hat{R}_{jk}  \right\}_{j,k=1,j\neq k}^{N}. 
	\end{equation} 
	\vspace{-0.6cm}
	
	\paragraph*{Distance measure}
	
	Following \cite{Buckwar2019}, we use the integrated absolute error (\textrm{IAE}) as distance between two functions $g_1,g_2$, 
	given by
	\begin{equation*}
		\textrm{IAE}(g_1,g_2):=\int\limits_{\mathbb{R}} | g_1(x)-g_2(x) | \ dx \ \in \ \mathbb{R}^+,
	\end{equation*}
	which can be approximated by rectangular integration. Then, the distance between the summaries $s(y)=\{ \hat{f}_{k}, \hat{S}_{k}, \hat{R}_{jk}\}$ of the observed dataset $y$ and the summaries $s(\tilde{y}_\theta)=\{ \tilde{f}_{k}, \tilde{S}_{k}, \tilde{R}_{jk}\}$ of a synthetic dataset $\tilde{y}_\theta$ (simulated via Algorithm \ref{Algorithm_Splitting}) is defined as
	\begin{eqnarray}
		\nonumber
		D(s(y),s(\tilde{y}_\theta))&:=& v_1 \frac{ 1}{N} \sum\limits_{k=1}^{N} \textrm{IAE}(\hat{S}_{k},\tilde{S}_{k})  
		\label{eq:ABC_dist} +v_2\frac{ 1}{N} \sum\limits_{k=1}^{N} \textrm{IAE} (\hat{f}_{k},\tilde{f}_{k}) \\ &&   
		+v_3 \frac{ 1}{N(N-1)} \sum\limits_{j,k=1,j\neq k}^{N}  \textrm{IAE}(\hat{R}_{jk},\tilde{R}_{jk}).
	\end{eqnarray} 
	The values $v_1,v_2,v_3$ are weights guaranteeing that the different summary functions have a comparable impact on the distance measure (cf. Section \ref{sec:Impl_det}).
	
	
	\subsubsection{Algorithm implementation details}
	\label{sec:Impl_det}
	
	The nSMC-ABC method is coded using the statistical software \texttt{R} \cite{R}, combined with the \texttt{Rcpp} package \cite{Rcpp}, offering a seamless integration of \texttt{R} and \texttt{C++}. In particular, the splitting simulation (see Algorithm~\ref{Algorithm_Splitting} and Appendix \ref{app:splitting}) is coded in \texttt{C++} (with an \texttt{R}-package provided at \url{https://github.com/massimilianotamborrino/StrangSplittingJRNMM}) and then integrated into the \texttt{R}-code for the nSMC-ABC Algorithm~\ref{alg:SMC_SBP_ABC}. 
	The for-loops of Algorithm~\ref{alg:SMC_SBP_ABC} are parallelized using the packages \texttt{doParallel} and \texttt{foreach}. All experiments are run on multiple core High-Performance-Clusters. 
	
	The summaries \eqref{eq:ABC_summaries} are computed as follows: Estimates of the spectral densities $\hat{S}_k$, the densities $\hat{f}_k$ and the cross-correlation functions $\hat{R}_{jk}$ are obtained using the smoothed periodogram estimator \texttt{spectrum}, the (Gaussian) kernel density estimator $\texttt{density}$ and the \texttt{R}-function \texttt{ccf}, respectively. The weights $v_1, v_2, v_3$ in the distance function \eqref{eq:ABC_dist} are obtained as follows: We set $v_1=1$ and obtain $v_2$ and $v_3$ by dividing the average area below the spectral densities of the observed data by the average area below the densities (equal to $1$) and cross-correlation functions of the observed data, respectively.
	
	A continuous particle is perturbed via the Gaussian kernel, i.e., $\theta_c^*\sim \mathcal{N}(\theta_c,\hat\Sigma_{c,r})$, using the \texttt{R}-function \texttt{rmvn} (line $18$ of Algorithm \ref{alg:SMC_SBP_ABC}), with normal density computed with the \texttt{R}-function \texttt{dmvn} (line~$26$ of Algorithm~\ref{alg:SMC_SBP_ABC}) of the \texttt{R}-package \texttt{mvnfast}, providing computationally efficient tools for the multivariate normal distribution. 
	The probability $q_{\textrm{stay}}$ of the Bernoulli type perturbation kernel \eqref{kernel_d2} is set to $0.9$ (except for Appendix \ref{app:perturbationKernel}, where we investigate different~choices). 
	
	The number $M$ of kept particles per iteration is set to $500$. The initial threshold $\delta_1$ for the first iteration is obtained by a reference table acceptance-rejection ABC pilot run. Under the given prior, we produce $10^4$ distances and then choose $\delta_1$ as their median. For $r>1$, the threshold $\delta_r$ is chosen as the median of the $M$  distances computed at the previous iteration if the acceptance rate of particles at the previous iteration is larger than 1\%. Otherwise $\delta_r$ is chosen as the $75$-th percentile  
	of the $M$ distances computed at iteration $r-1$ (to enable further iterations of the algorithm at its final stage). The algorithm is stopped after the acceptance rate has dropped below the prefixed threshold of 0.1\%. 
	
	Sample code is provided at {\small{\url{https://github.com/IreneTubikanec/networkABC}}}, with a comprehensive R-package at \url{https://github.com/massimilianotamborrino/SMCABCnJRNMM}.

	
	\section{Network inference in the stochastic multi-population JR-NMM from simulated data}
	\label{sec:5}
	
	In this section, we test the performance of the proposed nSMC-ABC Algorithm \ref{alg:SMC_SBP_ABC} on simulated datasets. 
	
	
	\subsection{Parameter vector and prior distribution}
	\label{sec:5:1}
	
	We aim to infer both the continuous and the binary parameters of the stochastic multi-population JR-NMM. The continuous parameters consist of the intrinsic model parameters $A_k>0$, $k=1,\ldots,N$, which play a central role in the (non-)activation of neural populations (cf. Section \ref{sec:2}), as well as the coupling strength parameters $L>0$ and $0 \ll c < 1$, see \eqref{eq:coupling} and \eqref{eq:coupling_structure}. The constants $\mu_k$ and $\sigma_k$ are fixed to $90$ and $500$, respectively, and the remaining continuous constants are fixed according to the typical values reported in Table~\ref{table1}. 
	The binary parameters consist of the coupling direction parameters $\rho_{jk}$, $j,k=1,\ldots,N$, $j\neq k$, see \eqref{eq:coupling}. Thus, our goal is to apply the nSMC-ABC Algorithm~\ref{alg:SMC_SBP_ABC} for inference of the $\text{$(N+2+N(N-1))$}$-dimensional parameter vector
	\begin{equation}\label{eq:theta_sim}
		\theta=(\underbrace{A_1,\ldots,A_N,L,c}_{\theta_c},\underbrace{\textrm{vec}(\mathcal{P})}_{\theta_b}),
	\end{equation}
	where the binary parameters $\theta_b=\textrm{vec}(\mathcal{P})$ are given by
	\begin{equation}\label{eq:P}
		\mathcal{P}=\begin{pmatrix}
			- & \rho_{12} & \ldots & \ldots & \rho_{1N} \\
			\rho_{21} & -  & \ddots &  & \vdots  \\
			\vdots & \ddots & \ddots & \ddots & \vdots \\
			\vdots &   &  \ddots & - & \rho_{N-1N} \\
			\rho_{N1} & \ldots & \ldots & \rho_{NN-1} & - \\
		\end{pmatrix},
	\end{equation}
	with $\rho_{jk}\in\{0,1\}, j,k=1,\ldots, N, j\neq k.$ 
	
	As prior distributions for $\theta_c$ in \eqref{eq:theta_sim}, we use continuous uniforms with fixed supports, i.e., 
	\begin{equation*}
		A_k \sim \textrm{U}(2,4), \ k=1,\ldots,N, \qquad  L\sim \textrm{U}(100,2000), \qquad  c\sim \textrm{U}(0.5,1).
	\end{equation*}
	As prior distributions for $\theta_b$ in \eqref{eq:theta_sim}, we consider Bernoulli distributions with equal probabilities, i.e.,
	\begin{equation}
		\label{eq:rhopriors}
		\rho_{jk}\sim \textrm{Bernoulli}\left(p\right), \quad p=\frac{1}{2}, \quad j,k= 1,\ldots,N, \ j\neq k.
	\end{equation}
	
	
	\begin{figure}
		\centering
		\subfigure[]{\includegraphics[width=0.315\textwidth]{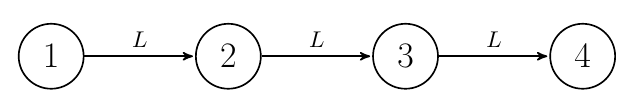}}
		\hspace{-0.2cm}
		\subfigure[]{\includegraphics[width=0.345\textwidth]{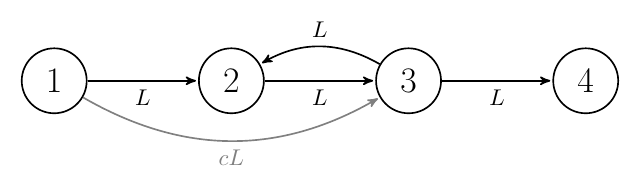}} 
		\hspace{-0.4cm}
		\subfigure[]{\includegraphics[width=0.345\textwidth]{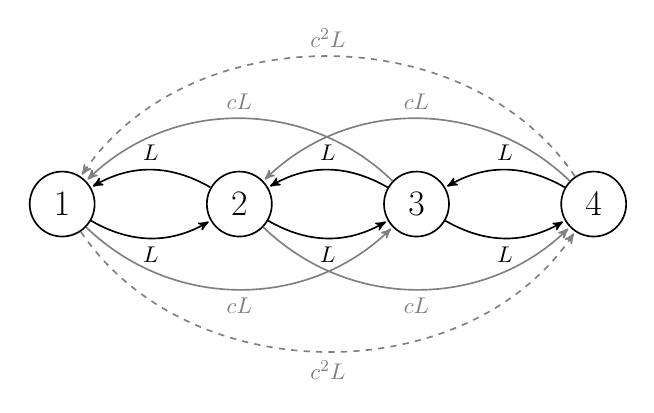}}
		\caption{{\bf Network structures} (under coupling strength \eqref{eq:coupling_structure}) of $N=4$ neural populations used to generate simulated observed data. (a)~Cascade network structure with equal coupling strengths. 
			(b) Partially connected network. (c) Fully connected network. 
		}
		\label{fig:network_structures_N4}
	\end{figure}
	
	\subsection{Description of observed data}
	
	We focus on $N=4$ neural populations and consider three network scenarios (under the coupling strength structure \eqref{eq:coupling_structure}): (a) a cascade network (as used to simulate the data in Figure \ref{fig:paths_coupling}), (b) a partially connected network and (c) a fully connected network. These three settings are visualized in Figure \ref{fig:network_structures_N4}. 
	
	Under each setting, we obtain a different multivariate time series reference dataset $y$. Specifically, the three resulting reference datasets $y$ are generated (via Algorithm \ref{Algorithm_Splitting}) up to time $T=20$ with time step $10^{-4}$. They are then subsampled with observation time step $\Delta=2 \cdot 10^{-3}$, yielding a $4$-dimensional time series dataset $y$ with $4$ components $y^k=(y^k(t_i))_{i=0}^{m}$, $k=1,\ldots,4$, of length $m=10^4$ each. The underlying parameter values are 	
	\begin{align}\label{eq:thetaTrue_sim}
		\begin{split}
			\theta&=(A_1,A_2,A_3,A_4,L,\textrm{vec}(\mathcal{P}))=(3.6,3.25,3.25,3.25,700,\textrm{vec}(\mathcal{P})),\\ & \ \text{with } \rho_{12}=\rho_{23}=\rho_{34}=1 \text{ and all other } \rho_{jk}=0 \mbox{ (cascade)},\\
			\theta&=(A_1,A_2,A_3,A_4,L,c,\textrm{vec}(\mathcal{P}))=(3.6,3.25,3.25,3.25,700,0.8,\textrm{vec}(\mathcal{P})),\\ & \ \text{with } \rho_{12}=\rho_{23}=\rho_{34}=\rho_{13}=\rho_{32}=1 \text{ and all other } \rho_{jk}=0 \mbox{ (partially connected)},\\
			\theta&=(A_1,A_2,A_3,A_4,L,c,\textrm{vec}(\mathcal{P}))=(3.25,3.25,3.25,3.25,700,0.8,\textrm{vec}(\mathcal{P})),\\ & \ \text{with } \rho_{jk}=1 \text{ for all } j,k \mbox{ (fully connected)}.
		\end{split}
	\end{align}
	In the cascade scenario, the parameter $c$ is not present and is excluded from $\theta$. 
	
	
	\subsection{Estimation results} 
	\label{sec:5:2}

	
	\begin{figure}
		\begin{centering}
			\includegraphics[width=1.0\textwidth]{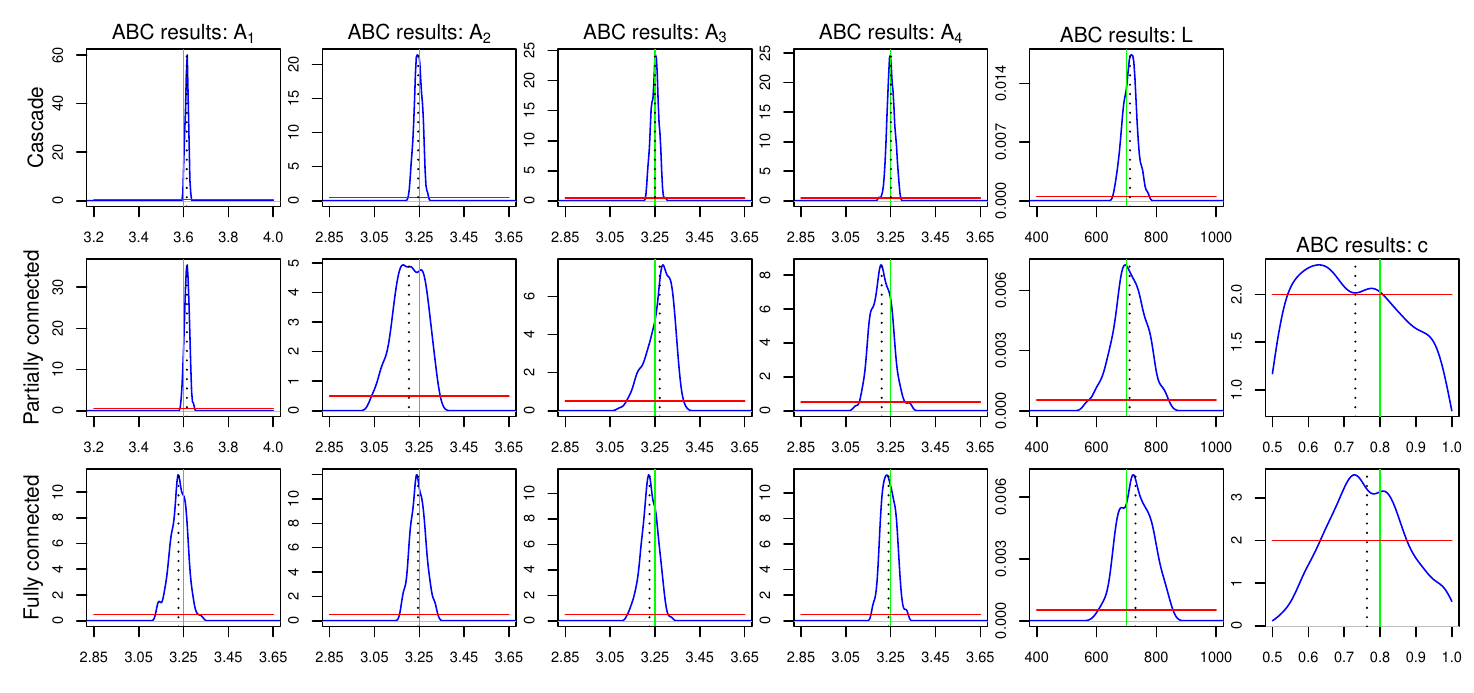}	
			\caption{\textbf{nSMC-ABC marginal posterior densities} (blue lines) compared to the prior densities (horizontal red lines) of the continuous parameters in the cascade (top panels), partially connected (middle panels) and fully connected (bottom panels) scenario, respectively. The vertical green lines and the dotted black lines indicate the true parameter values and the weighted marginal posterior means, respectively.}
			\label{fig:ABC_results_cont_sim}
		\end{centering}
	\end{figure}
	
	We apply the nSMC-ABC Algorithm~\ref{alg:SMC_SBP_ABC} to the three reference datasets $y$ described above in order to estimate $\theta$ \eqref{eq:theta_sim}. Figure \ref{fig:ABC_results_cont_sim} shows the marginal posterior densities (blue lines) and the uniform prior densities (horizontal red lines) of the continuous parameters for the cascade (top panels), partially connected (middle panels) and fully connected (bottom panels) network, respectively. The corresponding true parameter values \eqref{eq:thetaTrue_sim} used to generate the observed data are indicated by vertical solid green lines, while the dotted black lines are the respective weighted marginal posterior means. They are given by
	\begin{align*}
		(\hat{A}_1,\hat{A}_2, \hat{A}_3, \hat{A}_4, \hat{L})&=(3.614,3.245,3.250,3.251,711.986),\\
		(\hat{A}_1,\hat{A}_2, \hat{A}_3, \hat{A}_4, \hat{L}, \hat{c})&=(3.615,3.204,3.271,3.210,710.707,0.731),\\
		(\hat{A}_1,\hat{A}_2, \hat{A}_3, \hat{A}_4, \hat{L}, \hat{c})&=(3.227,3.244,3.226,3.240,730.328,0.764),
	\end{align*}
	for the three scenarios, respectively, and closely resemble the corresponding true values in \eqref{eq:thetaTrue_sim}. In all three scenarios, we obtain unimodal and narrow marginal posterior densities covering the true values for $L$ and $A_k$, $k=1,\ldots,4$. While the inference for the parameter $c$ is not satisfactory in the partially connected scenario (possibly due to the low information about $c$ available in the data, as it only enters through the connection from Population $1$ to Population~$3$, cf. Figure \ref{fig:network_structures_N4}b), for the fully connected network we observe a clear update of the approximate posterior compared to the prior also for that parameter.
	
	In Table \ref{table:ABC_results_rho_sim}, we report the estimates $\hat\rho_{jk}$ of $\rho_{jk}$  obtained as ABC posterior modes, together with the ABC posterior means in parentheses. All estimates coincide with the underlying true values of the $\rho_{jk}$ parameters in \eqref{eq:thetaTrue_sim}, the full network thus being correctly identified for all scenarios. In most cases (except for $\rho_{13}$ and $\rho_{14}$ in the partially connected scenario), even the posterior means coincide with the true values for $\rho_{jk}$ being equal to $0$ or $1$, respectively.
	
	\setlength{\tabcolsep}{4pt}
	\begin{table}[t]
		{  
			\caption{nSMC-ABC network estimates of ${\rho}_{jk}$ obtained as marginal posterior modes (and corresponding marginal posterior means in parentheses) for the cascade, partially connected and fully connected network scenarios, respectively. 
			}
			\vspace{-0.5cm}
			\label{table:ABC_results_rho_sim}
			\begin{center}
				\scalebox{0.85}{
					\begin{tabular}{l|lllllllllllll}
						\hline 
						Scenario & $\hat\rho_{12}$ & $\hat\rho_{13}$ & $\hat\rho_{14}$ & $\hat\rho_{21}$ & $\hat\rho_{23}$ & $\hat\rho_{24}$ & $\hat\rho_{31}$ & $\hat\rho_{32}$ & $\hat\rho_{34}$ & $\hat\rho_{41}$ & $\hat\rho_{42}$ & $\hat\rho_{43}$ \\  
						\hline
						Cascade & 1 (1) & 0 (0) & 0 (0) & 0 (0) & 1 (1) & 0 (0) & 0 (0) & 0 (0) & 1 (1) & 0 (0) & 0 (0) & 0 (0) \\	
						Partially connected & 1 (1) & 1 (0.99) & 0 (0.04) & 0 (0) & 1 (1) & 0 (0) & 0 (0) & 1 (1) & 1 (1) & 0 (0) & 0 (0) & 0 (0) \\	
						Fully connected & 1 (1) & 1 (1) & 1 (1) & 1 (1) & 1 (1) & 1 (1) & 1 (1) & 1 (1) & 1 (1) & 1 (1) & 1 (1) & 1 (1) \\	
						\hline
				\end{tabular}}
		\end{center}}
	\vspace{-0.4cm}
	\end{table}  
	
	
	\section{Network inference in the stochastic multi-population JR-NMM from EEG data with epileptic activity}\label{sec:6}
	\vspace{-0.2cm}
	
	After validating the proposed nSMS-ABC Algorithm \ref{alg:SMC_SBP_ABC} on simulated data, we now use it to estimate the connectivity structure from real EEG data. 
	
	
	\vspace{-0.3cm}
	\subsection{Description of the data}
	\label{sec:6:1}
	\vspace{-0.1cm}
	
	The investigated data is taken from the \textit{CHB-MIT Scalp EEG Database} \cite{Guttag2010} available on \textit{PhysioNet} \cite{Goldberger2000} at {\small{\url{https://www.physionet.org/content/chbmit/1.0.0/}}}. 
	This database has been collected at the Children's Hospital Boston and contains a set of EEG recordings from 22 pediatric subjects with intractable seizures, which were monitored for up to several days. In \cite{Shoeb2009}  seizure periods were detected within these datasets. 
	
	We analyze two $80$ seconds recordings of the first pediatric subject, an $11$~year old female patient, available in the edf-files {\tt chb01\_03} and {\tt chb01\_15} and denoted by {\tt data1} and {\tt data2}, respectively. In Appendix \ref{app:otherEEGpatient}, two recordings from the second patient are analyzed. Following \cite{Ostergaard2017} and \cite{Ruse2019}, we consider the four channels FP1-F7, FP1-F3, FP2-F4 and FP2-F8, where FP refers to the frontal lobes (the first two on the left brain hemisphere and the second two on the right) and F to a set of electrodes placed behind them. The electrode locations are according to the international 10--20 system for EEG measurements. The data are visualized in Figure~\ref{fig:EEG_data_chb01}, where the vertical dotted red lines separate the data into the period before and during seizure, lasting $40$ seconds each. The data are sampled at $256$~Hz, corresponding to $10240$ discrete time measurements during a $40$ seconds period. To put the recordings on the same scale as the model, we multiply each data point by $0.05$.

	
	\vspace{-0.2cm}
	\subsection{Inference}
	\label{sec:6:2}
	\vspace{-0.1cm}
	
\begin{figure}[H]
		\centering
		\subfigure[{\tt data1}: before and during seizure]{\includegraphics[width=0.95\textwidth]{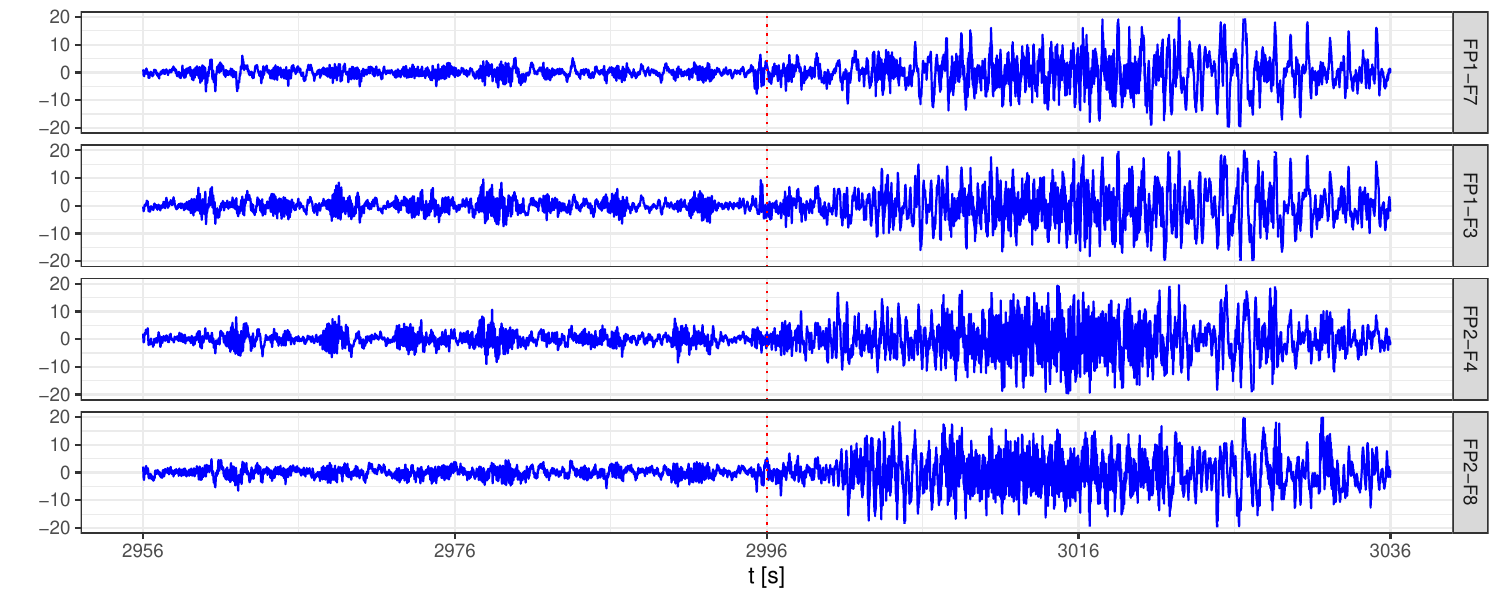}}
		\subfigure[{\tt data2}: before and during seizure]{\includegraphics[width=0.95\textwidth]{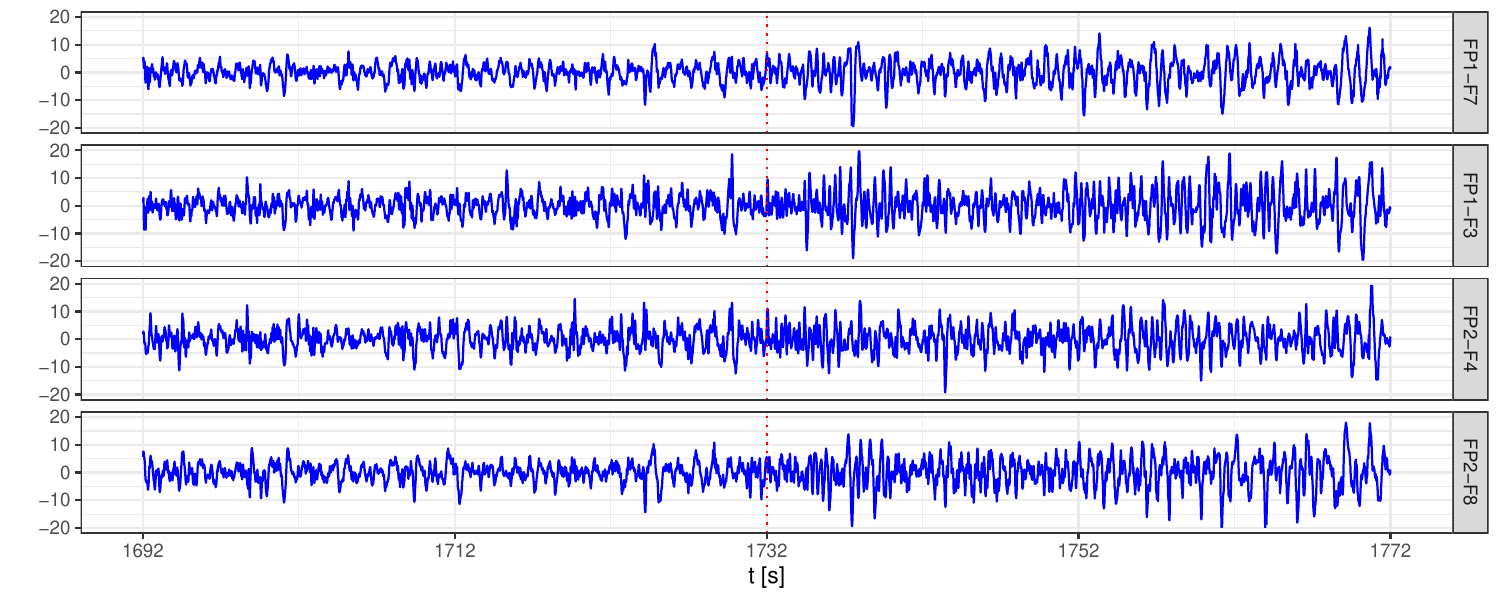}} 
		\vspace{-0.2cm}
		\caption{{\bf Two EEG recordings of an $11$ year old female patient.} The intervals $[2996,3036]s$ in {\tt data1} (a) and  $[1732,1772]s$ in {\tt data2} (b) (measurements to the right of the vertical dotted red lines) have been classified as  seizures in \cite{Shoeb2009}.} 
		\label{fig:EEG_data_chb01}
	\end{figure}

	We now use the nSMC-ABC Algorithm~\ref{alg:SMC_SBP_ABC} to fit the stochastic multi-population JR-NMM to the four multivariate EEG segments shown in Figure~\ref{fig:EEG_data_chb01} (i.e., {\tt data1} before seizure, {\tt data1} during seizure, {\tt data2} before seizure, and {\tt data2} during seizure). We infer the underlying network structure as well as relevant continuous model parameters in these four regimes. Our results offer possible insights into brain behavior when a seizure occurs.
	
	\vspace{-0.2cm}
	\paragraph*{Parameter vector and prior distribution}
	
	Denote the channels FP1-F7, FP1-F3, FP2-F4 and FP2-F8 by Population $1$, $2$, $3$ and $4$, respectively. As the distance between channels on different hemispheres is larger than that on the same hemisphere, the distance between the channels FP1-F3 and FP2-F4 is larger than that between FP1-F7 and FP1-F3 or between FP2-F4 and FP2-F8. Similar to \eqref{eq:coupling_structure}, we therefore assume the following matrix of coupling strength parameters
	\begin{equation}\label{eq:mat_CouplingStrengthStructure}
		K=\begin{pmatrix}
			- & K_{12} & K_{13} & K_{14}  \\
			K_{21} & -  & K_{23} & K_{24}  \\
			K_{31} & K_{32} & - & K_{34} \\
			K_{41} & K_{42} & K_{43} & -
		\end{pmatrix}
		=\begin{pmatrix}
			- & L & c^2L & c^3L  \\
			L & -  & cL & c^2L  \\
			c^2L & cL & - & L \\
			c^3L & c^2L & L & -
		\end{pmatrix},
	\end{equation}
	with unknown parameters $L>0$ and $0 \ll c<1$.

	\begin{remark}
		We verified that different choices of the coupling strength structure \eqref{eq:mat_CouplingStrengthStructure} lead to similar inferential results, in particular to the same network estimates (cf. Appendix \ref{App:CouplingStrengthStructure} for details). 
	\end{remark}
	
	We assume the continuous model parameters, i.e., the activation parameters $A_k$, $k=1,\ldots,4$, the noise intensity parameters $\sigma_L:=\sigma_1=\sigma_2$ (left hemisphere) and $\sigma_R:=\sigma_3=\sigma_4$ (right hemisphere) as well as the input parameters $\mu_L:=\mu_1=\mu_2$ (left hemisphere) and $\mu_R:=\mu_3=\mu_4$ (right hemisphere) to be unknown. For the remaining parameters, we use the typical values reported in Table \ref{table1}, except for setting $b=20$ and $C=70$ (values chosen based on pilot experiments). We thus aim to infer the $(10+12)$-dimensional parameter vector
	\begin{equation*}
		\theta=(\underbrace{A_1,A_2,A_3,A_4,L,c,\sigma_L,\sigma_R,\mu_L,\mu_R}_{\theta_c},\underbrace{\textrm{vec}(\mathcal{P})}_{\theta_b}), 
	\end{equation*}
	with $\mathcal{P}$ as in \eqref{eq:P} for $N=4$. We choose continuous uniform priors for $\theta_c$ with broad supports, 
	\begin{gather*}
		A_k \sim \textrm{U}(1,15), \ k=1,\ldots,4, \quad  L\sim \textrm{U}(100,3000), \quad  c\sim \textrm{U}(0.5,1),\\
		\sigma_L,\sigma_R \sim \textrm{U}(100,15000), \quad \mu_L,\mu_R \sim \textrm{U}(1,200).
	\end{gather*}
	The priors for $\theta_b$ are Bernoulli distributions with equal probabilities, as in \eqref{eq:rhopriors} for $N=4$.

	\paragraph*{Estimation results} 
	
	\begin{figure}[t]
		\begin{centering}
			\includegraphics[width=1.0\textwidth]{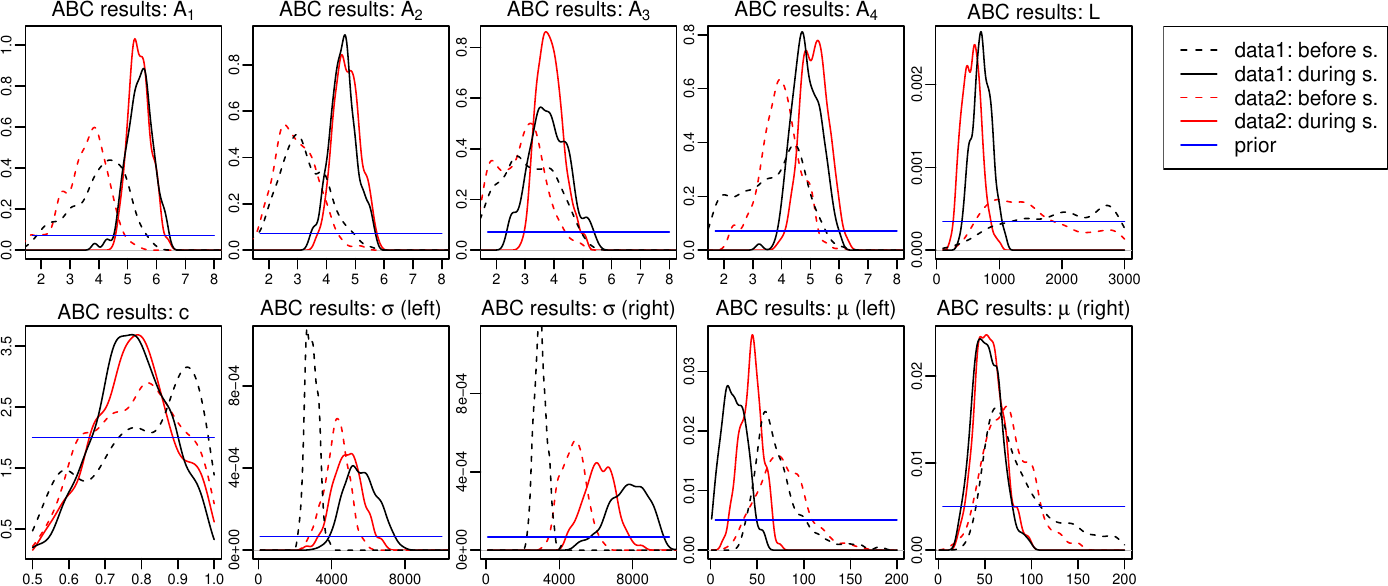}	
			\caption{\textbf{nSMC-ABC marginal  
					posterior densities} of the continuous parameters before (dashed lines) and during (solid lines) seizure ({\tt data1}: black, {\tt data2}: red). The horizontal blue lines are the respective uniform prior densities.}
			\label{fig:ABC_results_EEG_cont}
		\end{centering}
	\end{figure}

	In Figure \ref{fig:ABC_results_EEG_cont}, we report the marginal posterior densities for {\tt data1} (black lines) and {\tt data2} (red lines) and the corresponding uniform prior densities (horizontal blue lines) of the continuous parameters $\theta_c^u$, $u=1,\ldots, 10$. The results for the corresponding before and during seizure scenarios are indicated by dashed and solid lines, respectively. All posteriors for the during seizure scenarios and most posteriors for the before seizure scenarios are unimodal and show a clear update compared to the priors. We observe a similar behavior within the two before (resp. during) seizure scenarios as well as clear differences across the before and during seizure scenarios. The results for both datasets suggest a larger activation in all $4$ populations during seizure, as the posterior supports of  parameters $A_k$, $k=1,\ldots,4$, are shifted to the right. As expected, the posterior supports of the noise parameters $\sigma_L$ and $\sigma_R$ are also shifted towards larger values during seizures, implying larger variance in both hemispheres during seizure. This right shift is clearer for the first dataset. In contrast, the posteriors for $\mu_L$ and $\mu_R$ are shifted towards smaller values during seizure. While we obtain clear posterior estimates of $L$ during seizure, this is not the case before seizure. However, in all scenarios the posteriors  suggest a coupling strength clearly away from zero.
	
			\setlength{\tabcolsep}{1pt}
	\begin{table}
		{  
			\caption{nSMC-ABC network estimates of ${\rho}_{jk}$ obtained as marginal posterior modes (marginal posterior means in parentheses) for the before and during seizure periods of {\tt data1} and~{\tt data2}.}
			\vspace{-0.5cm}
			\label{table:ABC_results_rho_eeg_chb01}
			\begin{center}
				\scalebox{0.8}{
					\begin{tabular}{l|llllllllllll}
						\hline 
						EEG data & $\hat\rho_{12}$ & $\hat\rho_{13}$ & $\hat\rho_{14}$ & $\hat\rho_{21}$ & $\hat\rho_{23}$ & $\hat\rho_{24}$ & $\hat\rho_{31}$ & $\hat\rho_{32}$ & $\hat\rho_{34}$ & $\hat\rho_{41}$ & $\hat\rho_{42}$ & $\hat\rho_{43}$ \\  
						\hline
						{\tt data1}: b.s. & 1 (0.842) & 1 (0.678) & 1 (0.896) & 0 (0.060) & 1 (0.810) & 0 (0.108) & 0 (0.226) & 1 (0.852) & 0 (0.058) & 1 (0.922) & 1 (0.512) & 1 (0.626) \\	
						{\tt data1}: d.s. & 1 (1) & 1 (0.786) & 1 (0.998) & 1 (1) & 1 (0.984) & 0 (0.004) & 0 (0.006) & 1 (0.994) & 0 (0.004) & 1 (0.998) & 0 (0.038) & 0 (0.006) \\	
						\hline
						{\tt data2}: b.s. & 1 (0.916) & 1 (0.856) & 1 (0.754) & 0 (0.030) & 1 (0.562) & 0 (0.086) & 1 (0.790) & 1 (0.814) & 1 (0.924) & 1 (0.958) & 1 (0.650) & 1 (0.996)  \\	
						{\tt data2}: d.s. & 1 (1) & 1 (0.972) & 0 (0.108) & 1 (1) & 1 (0.998) & 0 (0.002) & 0 (0.016) & 1 (0.948) & 1 (1) & 1 (0.996) & 0 (0.012) & 0 (0.070) \\	
						\hline
				\end{tabular}}
		\end{center}}
	\end{table}

\begin{figure}
		\centering
		\subfigure[Inferred network for {\tt data1}: before (left) and during (right) seizure]
		{\includegraphics[width=1.0\textwidth]{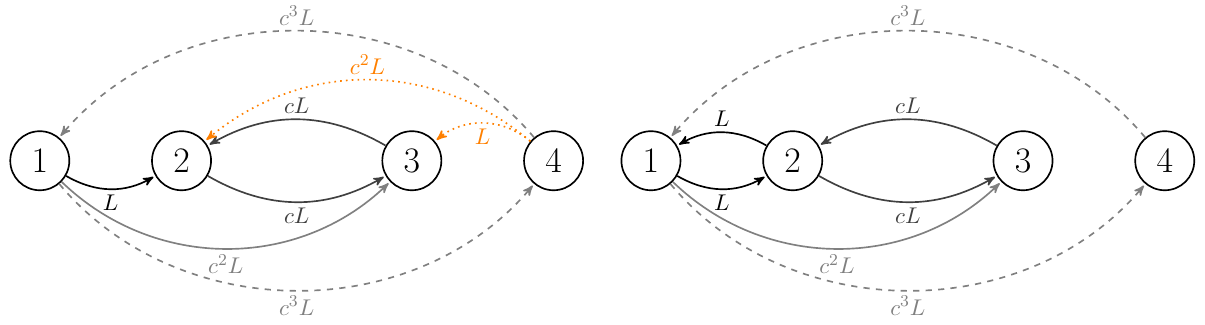}}
		\subfigure[Inferred network for {\tt data2}: before (left) and during (right) seizure]
		{\includegraphics[width=1.0\textwidth]{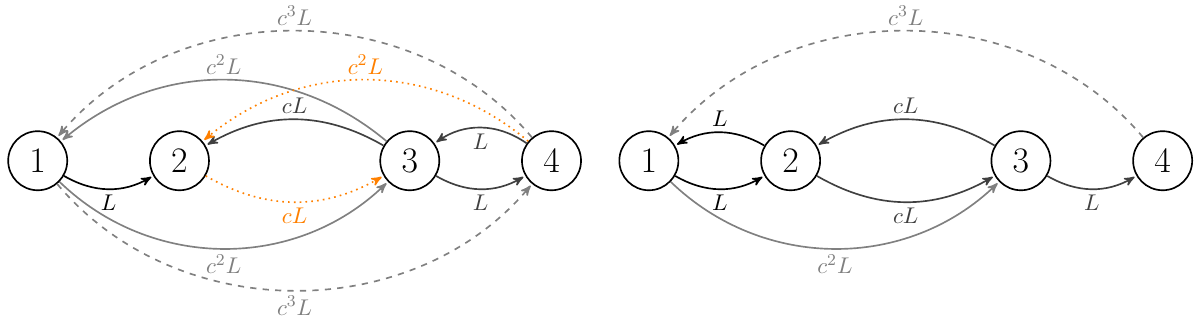}}
		\caption{{\bf Inferred networks from EEG recordings.} Estimated networks from two recordings with $N=4$ channels of an 11 year old female patient before and during epileptic seizure. Populations 1, 2, 3, and 4 refer to channels FP1-F7, FP1-F3, FP2-F4, and FP2-F8, respectively. The dotted orange connections are estimated with a posterior mean within $[1/3,2/3]$  
			(cf. Table \ref{table:ABC_results_rho_eeg_chb01}).  
		}
		\label{fig:ABC_results_EEG_net}
	\end{figure}

	In Table \ref{table:ABC_results_rho_eeg_chb01}, we report estimates (values in $\{0,1\}$ obtained as marginal posterior modes) for the $12$ network parameters $\rho_{jk}$, $j,k=1,\ldots,4$, $j\neq k$, as well as the corresponding posterior means in parentheses. All posterior means for the during seizure scenarios and most posterior means for the before seizure scenarios lie outside the interval $[1/3,2/3]$, being thus ``close'' to $0$ or $1$, respectively. Therefore, the results suggest clear connectivity structures, with the estimated networks visualized in Figure \ref{fig:ABC_results_EEG_net}. Remarkably, we observe similarities in the inferred networks for both datasets.  
	For example, the estimation results suggest a stronger connectivity in the left brain hemisphere during seizure, compared to before seizure (in particular, for both datasets the connection from Population 2 to Population 1 is only present during seizure). Moreover, for both datasets we observe less connectivity in the right brain hemisphere as well as less connectivity from the right to the left hemisphere during seizure, compared to before seizure. There are also differences among the two datasets though, e.g., the connection from Population 3 to Population~4 is only present for the second dataset.
	
	Overall, it is remarkable how the proposed nSMC-ABC method identifies clear similarities between the two datasets, despite their apparent differences (cf. Figure \ref{fig:EEG_data_chb01}). This may indicate that the functional network characterizes relevant differences between epileptic seizures and normal brain activity.

	\paragraph*{Fitted summaries}
	
\begin{figure}
		\begin{centering}
			\includegraphics[width=1.0\textwidth]{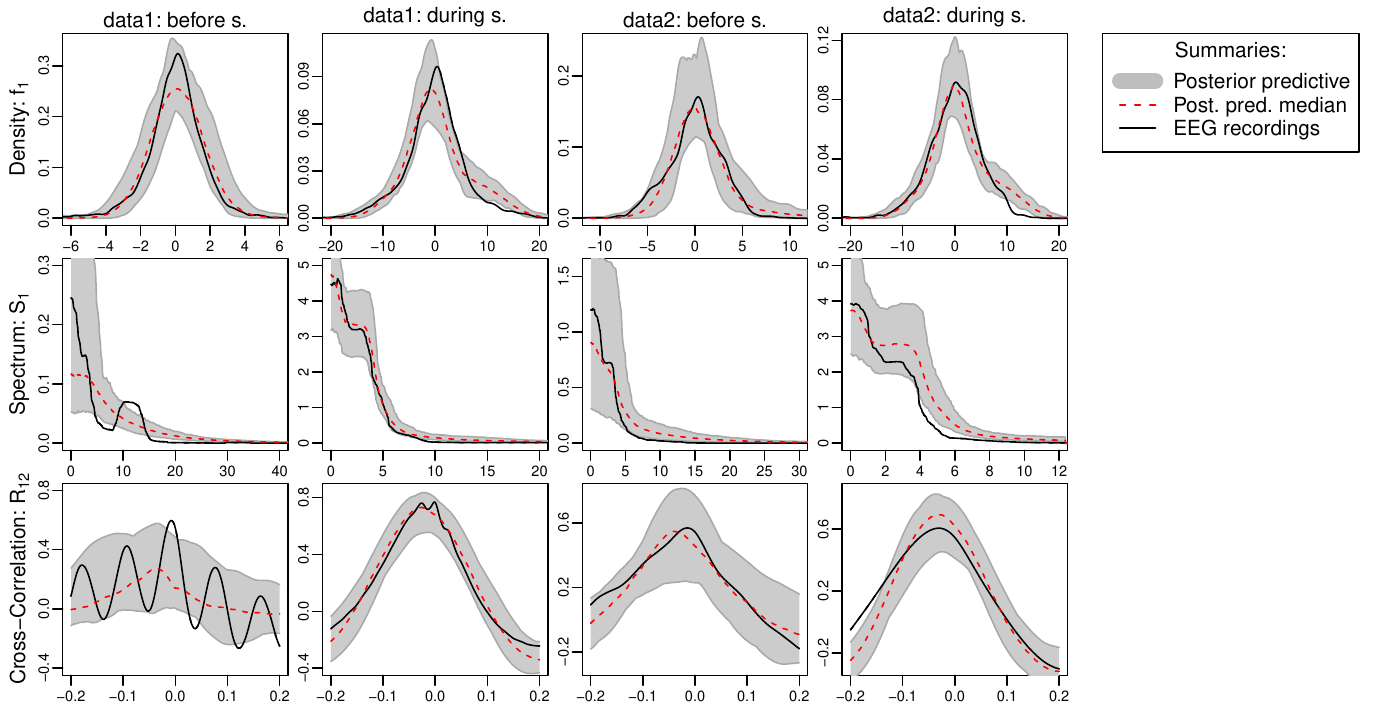}	
			\caption{\textbf{Summaries} $f_1$, $S_1$ and $R_{12}$ (cf. \eqref{eq:ABC_summaries}) of the EEG recordings (solid black lines) compared to summaries derived from synthetic datasets generated with $100$ kept posterior samples (gray bands and dashed red lines for their medians).}
			\label{fig:fittedSummaries_chb01}
		\end{centering}
	\end{figure}

	Figure \ref{fig:fittedSummaries_chb01} provides a comparison of the summary statistics \eqref{eq:ABC_summaries} estimated from the EEG recordings (solid black lines) and from the posterior predictive simulations (gray areas). The median of the posterior predictive bands is marked by dashed red lines. Only a subset of the summaries is shown, i.e., the density $f_1$, the spectral density $S_1$ and the cross-correlation function $R_{12}$, but similar results are obtained for all other summary functions. The left and middle left panels correspond to the summaries of the before and during seizure periods of the first dataset,   and the middle right and right panels to the summaries of the before and during seizure periods of the second dataset. The matches of the observed and the posterior predicted summaries are good for all datasets and periods. 
	
	\newpage
	
	\section{Conclusion and discussion}
	\label{sec:7}
	
	Understanding how different regions in the brain are connected and function together, particularly during epileptic seizures, is a problem of large interest in neuroscience. We contribute to this effort by proposing a statistical method for inference in a stochastic neural mass model that describes the activity of multiple coupled neural populations. Specifically, we extend the single-population stochastic JR-NMM \cite{Ableidinger2017} to an $N$-population model, resulting in a $6N$-dimensional SDE with a variety of continuous biologically motivated parameters as well as binary coupling direction parameters. The latter characterize the network structure among the $N$ populations and facilitate the parameter inference from multi-channel EEG recordings. A major challenge of this inferential task is that none of the $6N$ model coordinates are directly observable. Indeed, only the $N$-dimensional process \eqref{eqn:output}, a multivariate linear function of some coordinates, is measurable with EEG recording techniques. Moreover, due to the intricacy of the model, the underlying likelihood function is not available. 
	
	To address these challenges, we employ the likelihood-free ABC methodology, introducing the nSMC-ABC method, an adaptation of the gold-standard SMC-ABC algorithm. Our method leverages the introduced $\{0,1\}$-valued network indicators, allowing for a reduction of computationally expensive continuous parameters and enabling the use of an efficient perturbation kernel that independently updates continuous and binary parameters. Two key algorithm components are essential: a reliable and computationally efficient numerical simulation method for synthetic data generation, and informative summary statistics capturing model dynamics independent of the underlying intrinsic stochasticity. To this end, we further develop the ABC approach of \cite{Buckwar2019} to multiple interacting populations. First, we construct a numerical scheme based on the splitting approach. It exploits the model's Hamiltonian structure and scales automatically with $N$. Standard methods like Euler-Maruyama would either fail in estimation or make the algorithm computationally infeasible due to the requirement of very small time discretization steps (see \cite{Buckwar2019} for an illustration of this issue on the stochastic JR-NMM with $N=1$). Second, we propose summaries that incorporate both individual signal features and pairwise connectivity measures among signals by mapping the $N$-dimensional time series to the $N$ corresponding densities and spectral densities, and the $N(N-1)$ cross-correlation functions. 
	
	The nSMC-ABC method significantly reduces computational cost compared to standard SMC-ABC while achieving excellent estimation results. When applied to simulated data, it performs well in various scenarios, ranging from cascade connectivity structures to fully connected networks. Application to real EEG data provides distinct estimates of both continuous model parameters and network structures, revealing clear similarities in recordings from the same subject and marked differences between pre-seizure and seizure states. These findings may contribute to a better understanding of brain activities associated with epileptic seizures. Our real data results are further validated through posterior predictive checks, demonstrating a strong match between summary statistics derived from the experimental data and those generated using kept nSMC-ABC particles.
	
	The successful inference of the network parameters together with a relatively large number of continuous parameters in the case of $N=4$ populations is remarkable, considering that prior works on the single-population JR-NMM identified only up to four continuous parameters \cite{Buckwar2019,Rodriguesetal2021}.
	Finally, the proposed nSMC-ABC algorithm, along with our summaries, is directly applicable to other coupled ergodic SDEs for modeling stationary multivariate time series, provided that a suitable numerical simulation method is available. The accompanying \texttt{R}-package facilitates implementation.

	

	
	\newpage
	
	\appendix
	
	
	\section{Simulation of the stochastic multi-population JR-NMM: Numerical  splitting method}
	\label{app:splitting}

	In this section, we derive the numerical splitting scheme reported in Algorithm \ref{Algorithm_Splitting}. In Section \ref{app:splitting:1}, we rewrite the stochastic multi-population JR-NMM as a Hamiltonian type system. This clever and compact reformulation of the model constitutes the basis for the construction of the proposed splitting scheme, which we detail in Section \ref{app:splitting:2}. 
	
	
	\subsection{Formulation of the stochastic multi-population JR-NMM as a Hamiltonian type system}
	\label{app:splitting:1}
	
	The stochastic multi-population JR-NMM can be formulated as a damped stochastic Hamiltonian type system with non-linear displacement, similar as in \cite{Ableidinger2017} for $N=1$ neural population. 
	
	To do so, decompose the $k$-th process as $X^k:=(Q^k,P^k)^\top$ with $3$-dimensional components 
	$Q^k=(X_1^k,X_2^k,X_3^k)^\top$ and $P^k=(X_4^k,X_5^k,X_6^k)^\top$, for $k\in \{1,\ldots,N\}$. Denote the $3$-dimensional Wiener process in the $k$-th population by $W^k=(W_4^k,W_5^k,W_6^k)^\top$ and the corresponding $3\times 3$-dimensional diffusion matrix by $\Sigma_k=\text{diag}[\epsilon_{k},\sigma_{k},\epsilon_{k}]$. The Hamiltonian type formulation of the $k$-th population, described via system~\eqref{JRNMM_1Pop} with suitable index $k$ and equation \eqref{eq:coupling}, is then given by
	{\small{\begin{equation}\label{eqn:Hamiltonian:kPop}
				d \begin{pmatrix}
					Q^k(t) \\
					P^k(t) 
				\end{pmatrix}
				=
				\begin{pmatrix}
					\nabla_{P} H{_k}\bigl(Q^k(t),P^k(t)\bigr) \\
					-\nabla_{Q} H{_k}\bigl(Q^k(t),P^k(t)\bigr) 
					-2\Gamma_k P^k(t) + G_k({Q(t)})
				\end{pmatrix} dt \ + \
				\begin{pmatrix}
					\mathbb{O}_3 \\
					\Sigma_k
				\end{pmatrix} dW^k(t).
	\end{equation}}}\noindent
	System \eqref{eqn:Hamiltonian:kPop} consists of a \textit{Hamiltonian part} defined by the Hamiltonian function $\text{$H{_k}:\mathbb{R}^6 \to \mathbb{R}_0^+$}$ given by
	\begin{equation*}
		H{_k}(Q^k,P^k):=\frac{1}{2}\left( \norm{P^k}^2 + \norm{\Gamma_k Q^k}^2 \right),
	\end{equation*}
	with gradients $\nabla_{P} H{_k}\bigl(Q^k(t),P^k(t)\bigr)=P^k(t)$ and $\nabla_{Q} H{_k}\bigl(Q^k(t),P^k(t)\bigr)=\Gamma_k^2 Q^k(t)$, a \textit{damping part} determined by the $3\times 3$-dimensional diagonal matrix $\Gamma_k=\text{diag}[a_k,a_k,b_k]$, and a non-linear  \textit{displacement {and coupling} term} $G_k:\mathbb{R}^{3{N}}\to \mathbb{R}^3$ given by $G_k({Q(t)})$ as in \eqref{eq:G}, where $Q=(Q^1,\ldots,Q^N)^\top=(X_1^1,X_2^1,X_3^1,\ldots,X_1^N,X_2^N,X_3^N)^\top$. 
	
	\begin{remark} 
		Note that the function $G_k$ \eqref{eq:G}, which is used in Algorithm \ref{Algorithm_Splitting},  only depends on the $Q$-component of the system. Moreover, incorporating the coupling term of the $k$-th population (cf. \eqref{eq:coupling}) into the displacement function $G_k$ enables closed-form  expressions of all required components of the splitting framework, for an arbitrary number of populations $N$ (see Section~\ref{app:splitting:2} below). 
	\end{remark}
	
	To obtain a compact formulation of the stochastic $N$-population JR-NMM as a stochastic Hamiltonian type system, we define the process $X:=(Q,P)^\top$ with $Q$ as above, $3N$-dimensional component $P=(P^1,\ldots,P^N)^\top=(X_4^1,X_5^1,X_6^1,\ldots,X_4^N,X_5^N,X_6^N)^\top$
	and $3N$-dimensional Wiener process $W=(W^1,\ldots,W^N)^\top=(W_4^1,W_5^1,W_6^1,\ldots,W_4^N,W_5^N,W_6^N)^\top$. This yields the following $6N$-dimensional SDE
	\begin{equation}\label{eqn:Hamiltonian}
		d \begin{pmatrix}
			Q(t) \\
			P(t) 
		\end{pmatrix}
		=
		\begin{pmatrix}
			P(t) \\
			-\Gamma^2Q(t) -2\Gamma P(t) + G(Q(t))
		\end{pmatrix} dt \ + \
		\begin{pmatrix}
			\mathbb{O}_{3N} \\
			\Sigma
		\end{pmatrix} dW(t),
	\end{equation}
	where $\Gamma=\text{diag}[a_1,a_1,b_1,\ldots,a_N,a_N,b_N]$ and $\Sigma=\text{diag}[\epsilon_{1},\sigma_{1},\epsilon_1,\ldots,\epsilon_N,\sigma_{N},\epsilon_N]$ are $3N\times 3N$-dimensional diagonal matrices, and the displacement and coupling function $G:\mathbb{R}^{3N}\to \mathbb{R}^{3N}$ is given by $G(Q)=(G_1({Q}),\ldots,G_N({Q}))^\top$.
	
	
	\subsection{Construction of the numerical splitting method for the stochastic multi-population JR-NMM}
	\label{app:splitting:2}
	
	Let $0=t_0<\ldots<t_{m}=T$ be an equidistant partition of the  time interval $[0,T]$ with time steps $\Delta=t_{i+1}-t_i$ for $i=0,\ldots,m-1$, $m\in \mathbb{N}$. The numerical splitting approach consists of the following three steps \cite{Blanes2009,Mclachlan2002}:
	\begin{itemize}
		\item[(i)] Split the equation for the process $X(t)$ into $d\in \mathbb{N}$ explicitly solvable subequations for $X^{[z]}(t)$, $z=1,\ldots,d$;
		\item[(ii)] Derive the exact solution (flow) $\varphi_{\Delta}^{[z]}$ of the $z$-th subequation over an increment of length~$\Delta$, for $z=1,\ldots, d$, i.e., 
		for initial value $X^{[z]}(t_i)$, the $\Delta$-flow is $X^{[z]}(t_{i+1})=\varphi_{\Delta}^{[z]}(X^{[z]}(t_i))$;
		\item[(iii)] Compose the $d$ exact solutions in a suitable way. Prominent methods are the Lie-Trotter and Strang compositions, given by
		\begin{align*}
			\widetilde{X}^{\textrm{LT}}(t_{i+1})&=\left( \varphi_{\Delta}^{[1]} \circ \ldots \circ \varphi_{\Delta}^{[d]} \right)\left( \widetilde{X}^{\textrm{LT}}(t_{i}) \right),\\
			\widetilde{X}^{\textrm{S}}(t_{i+1})&=\left( \varphi_{\Delta/2}^{[1]} \circ \ldots \circ  \varphi_{\Delta/2}^{[d-1]} \circ  \varphi_{\Delta}^{[d]} \circ  \varphi_{\Delta/2}^{[d-1]} \circ \ldots \circ \varphi_{\Delta/2}^{[1]} \right)\left( \widetilde{X}^{\textrm{S}}(t_{i}) \right),
		\end{align*}
		respectively (where the order of the compositions can also be changed). In particular,  $\widetilde{X}^{\textrm{LT}}(t_{i})$ (resp. $\widetilde{X}^{\textrm{S}}(t_{i})$) is the Lie-Trotter (resp. Strang) approximation of $X(t_i)$ at time~$t_i$.
	\end{itemize}
	Strang compositions have been shown to outperform Lie-Trotter methods, yielding a better approximation of the true solution and preserving the model dynamics for larger time steps (see, e.g., \cite{Ableidinger2017,BuckwarTubikanec2022,Chevallier2020,pilipovic2022parameter,Tubikanec2022}). This is possibly due to the symmetry of the Strang splitting \cite{Ableidinger2017}, its smaller mean biases \cite{Tubikanec2022} and its higher-order one-step predictions \cite{pilipovic2022parameter}. We therefore apply the Strang splitting approach for system \eqref{eqn:Hamiltonian}, generalizing the method presented in \cite{Ableidinger2017}. 
	
	\paragraph*{Step (i): Choice of subequations}
	
	We separate the non-linear term $G(Q(t))$ of system~\eqref{eqn:Hamiltonian}, and consider $d=2$ subequations given by
	\begin{align}
		\label{eqn:Sub1}d \begin{pmatrix}
			Q^{[1]}(t) \\
			P^{[1]}(t) 
		\end{pmatrix}
		&=
		\begin{pmatrix}
			P^{[1]}(t) \\
			-\Gamma^2Q^{[1]}(t) -2\Gamma P^{[1]}(t) 
		\end{pmatrix} dt \ + \
		\begin{pmatrix}
			\mathbb{O}_{3N} \\
			\Sigma
		\end{pmatrix} dW(t), \\
		\label{eqn:Sub2}d \begin{pmatrix}
			Q^{[2]}(t) \\
			P^{[2]}(t) 
		\end{pmatrix}
		&=
		\begin{pmatrix}
			0_{3N} \\
			G(Q^{[2]}(t))
		\end{pmatrix} dt.
	\end{align}
	
	\paragraph*{Step (ii): Derivation of exact solutions}
	
	Write subequation \eqref{eqn:Sub1} (a linear SDE with additive noise) as 
	\begin{equation*}
		dX^{[1]}(t)=FX^{[1]}(t)dt+\Sigma_0 dW(t),
	\end{equation*}
	where
	\begin{equation*}
		F=\begin{pmatrix}
			\mathbb{O}_{3N} & \mathbb{I}_{3N}\\
			-\Gamma^2 & -2\Gamma
		\end{pmatrix}, \quad
		\Sigma_0=\begin{pmatrix}
			\mathbb{O}_{3N} \\
			\Sigma
		\end{pmatrix}.
	\end{equation*}
	
	\noindent Let $X^{[1]}(t_i)=(Q^{[1]}(t_i),P^{[1]}(t_i))^\top$ denote the solution of system \eqref{eqn:Sub1} at time $t_i$. The exact solution at time $t_{i+1}$ is then given by
	\begin{equation*}
		X^{[1]}(t_{i+1})=\varphi_\Delta^{[1]}\left( X^{[1]} (t_i) \right)=e^{F\Delta}X^{[1]}(t_i)+\xi_i(\Delta),
	\end{equation*}
	where $\xi_i(\Delta)$, $i=1,\ldots,m$, are independent $6N$-dimensional Gaussian random vectors with zero mean $\mathbb{E}[\xi_i(\Delta)]=0_{6N}$ and covariance matrix given by 
	\begin{equation*}
		\textrm{Cov}(\Delta)=\int\limits_{0}^{\Delta}e^{F(\Delta-s)}\Sigma_0\Sigma_0^\top \left( e^{F(\Delta-s)} \right)^\top ds.
	\end{equation*}
	For general matrices $F$, the exponential matrix $\textrm{Exp}(\Delta)=e^{F\Delta}$ and the covariance matrix $\textrm{Cov}(\Delta)$ may be costly to compute. However, due to the sparseness of $F$ coming from the damped Hamiltonian term, both $\textrm{Exp}(\Delta)$ and $\textrm{Cov}(\Delta)$ can be expressed in closed-form as in~\eqref{eq:ExpCov}, for an arbitrary number of populations $N$.
	
	Denote the solution of subequation \eqref{eqn:Sub2} (a non-linear ODE) at time $t_i$ by $X^{[2]}(t_i)=(Q^{[2]}(t_i),P^{[2]}(t_i))^\top$. Since the first component of the right side of system \eqref{eqn:Sub2} is zero and the second component only depends on $Q$, the exact solution at time $t_{i+1}$ is given~by
	\begin{equation*}
		X^{[2]}(t_{i+1})=\varphi_\Delta^{[2]}\left( X^{[2]} (t_i) \right)=X^{[2]}(t_i)+\Delta \begin{pmatrix}
			0_{3N} \\
			G(Q^{[2]}(t_i))
		\end{pmatrix}.
	\end{equation*}
	
	\paragraph*{Step (iii): Composition of exact solutions}
	
	The Strang splitting method for system \eqref{eqn:Hamiltonian} is then given~by
	\begin{equation}
		\label{eqn:S}\widetilde{X}(t_{i+1})=\left( \varphi_{\Delta/2}^{[2]} \circ  \varphi_{\Delta}^{[1]} \circ  \varphi_{\Delta/2}^{[2]} \right)\left( \widetilde{X}(t_{i}) \right).
	\end{equation}
	Using \eqref{eqn:S}, a path of the stochastic multi-population JR-NMM can then be simulated via Algorithm \ref{Algorithm_Splitting}.

	\begin{remark}
		An analogous analysis as in \cite{Ableidinger2017} yields that the derived splitting scheme \eqref{eqn:S} is convergent (in the strong mean-square sense) and preserves the qualitative behavior of the solution of \eqref{eqn:Hamiltonian}, e.g., amplitudes of oscillations,  marginal invariant densities and spectral densities. 
		Moreover, note that the Strang splitting method 
		\begin{equation*}
			\widetilde{X}(t_{i+1})=\left( \varphi_{\Delta/2}^{[1]} \circ  \varphi_{\Delta}^{[2]} \circ  \varphi_{\Delta/2}^{[1]} \right)\left( \widetilde{X}(t_{i}) \right)
		\end{equation*}
		is another possible choice. However, this composition requires to evaluate the more costly stochastic subsystem twice at every iteration step, generating twice as many pseudo-random numbers, which is why we refrain from using it.
	\end{remark}
	
	
	\section{Deeper investigation of the proposed \lowercase{n}SMC-ABC algorithm}
	\label{app:nSMCABC}
	
	
	\subsection{\lowercase{n}SMC-ABC: Advantage of binary network parameters}\label{app:binaryIndicators}
	
	In this section, we illustrate in detail the advantage of introducing binary parameters $\rho_{jk} \in \{ 0,1 \}$ \eqref{eq:coupling} for network inference via the proposed nSMC-ABC Algorithm \ref{alg:SMC_SBP_ABC}.
	
	Imagine that the binary coupling direction parameters $\rho_{jk} \in \{ 0,1 \}$ were absent in \eqref{eq:coupling}. Then, the continuous coupling strength parameters $K_{jk} \geq 0$ would have to be used to determine the network structure and a standard SMC-ABC algorithm would be required for their inference. 
	This would be computationally expensive, since $N(N-1)$ continuous network parameters would have to be inferred. Instead, the use of binary network indicators allows to reduce the number of continuous network parameters (here, to only two parameters $L$ and $c$, see \eqref{eq:coupling_structure}), which significantly reduces the number of model simulations required in nSMC-ABC to reach the desired posterior region, and thus the computational cost of the algorithm. In addition, for the standard SMC-ABC algorithm a criterion would be required to decide whether there is or not a connection from population $j$ to $k$ (e.g., the marginal posterior median of $K_{jk}$ being larger than some threshold value). By using binary indicators and nSMC-ABC, we avoid setting such decision criterion, which brings an extra level of arbitrariness. Therefore, using binary indicators and nSMC-ABC has two main advantages. First, it allows to reduce the number of continuous parameters, and thus the computational cost of the algorithm. Second, it does not require a  decision criterion for the presence or absence of network connections.
	
	In the following, we illustrate these two advantages, comparing the nSMC-ABC Algorithm~\ref{alg:SMC_SBP_ABC} (based on binary network parameters and two continuous coupling strength parameters) and the standard SMC-ABC algorithm (purely based on continuous network parameters) for Setting 2 ``Partially connected network'', see second equation in \eqref{eq:thetaTrue_sim} and Figure \ref{fig:network_structures_N4}b. Similar observations are made for the other settings. 
	
	\begin{figure}[H]
		\begin{centering}
			\includegraphics[width=1.0\textwidth]{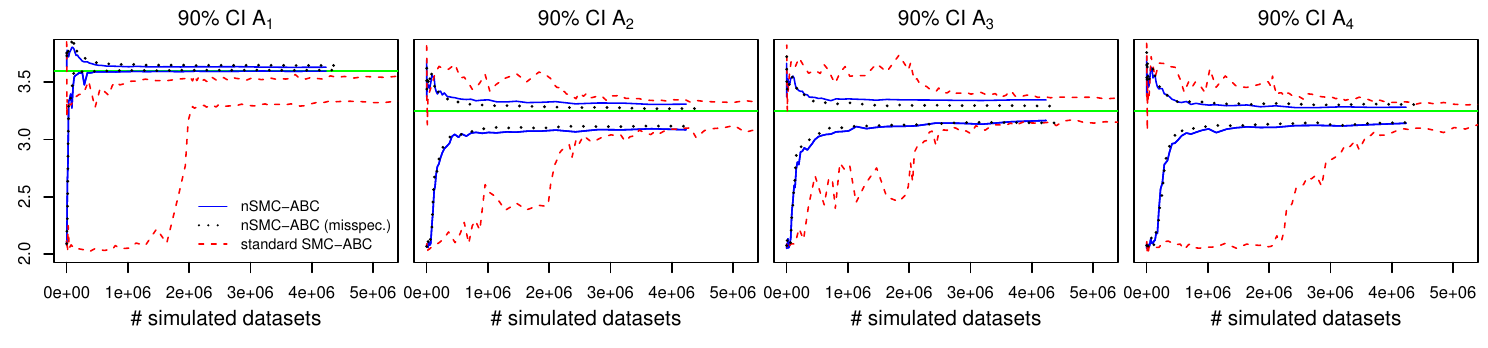}	
			\caption{\textbf{90\% CIs of the marginal posterior densities} of the $A$-parameters obtained from the nSMC-ABC Algorithm \ref{alg:SMC_SBP_ABC} (blue solid lines) and the standard SMC-ABC algorithm (red dashed lines), as a function of the number of model simulations. The green horizontal lines indicate the true parameter values. The black dotted lines correspond to the inferential results obtained via nSMC-ABC under a misspecified model scenario.}
			\label{fig:advantageBinaryAs}
		\end{centering}
	\end{figure}
	
	\begin{figure}[H]
		\begin{centering}
			\includegraphics[width=1.0\textwidth]{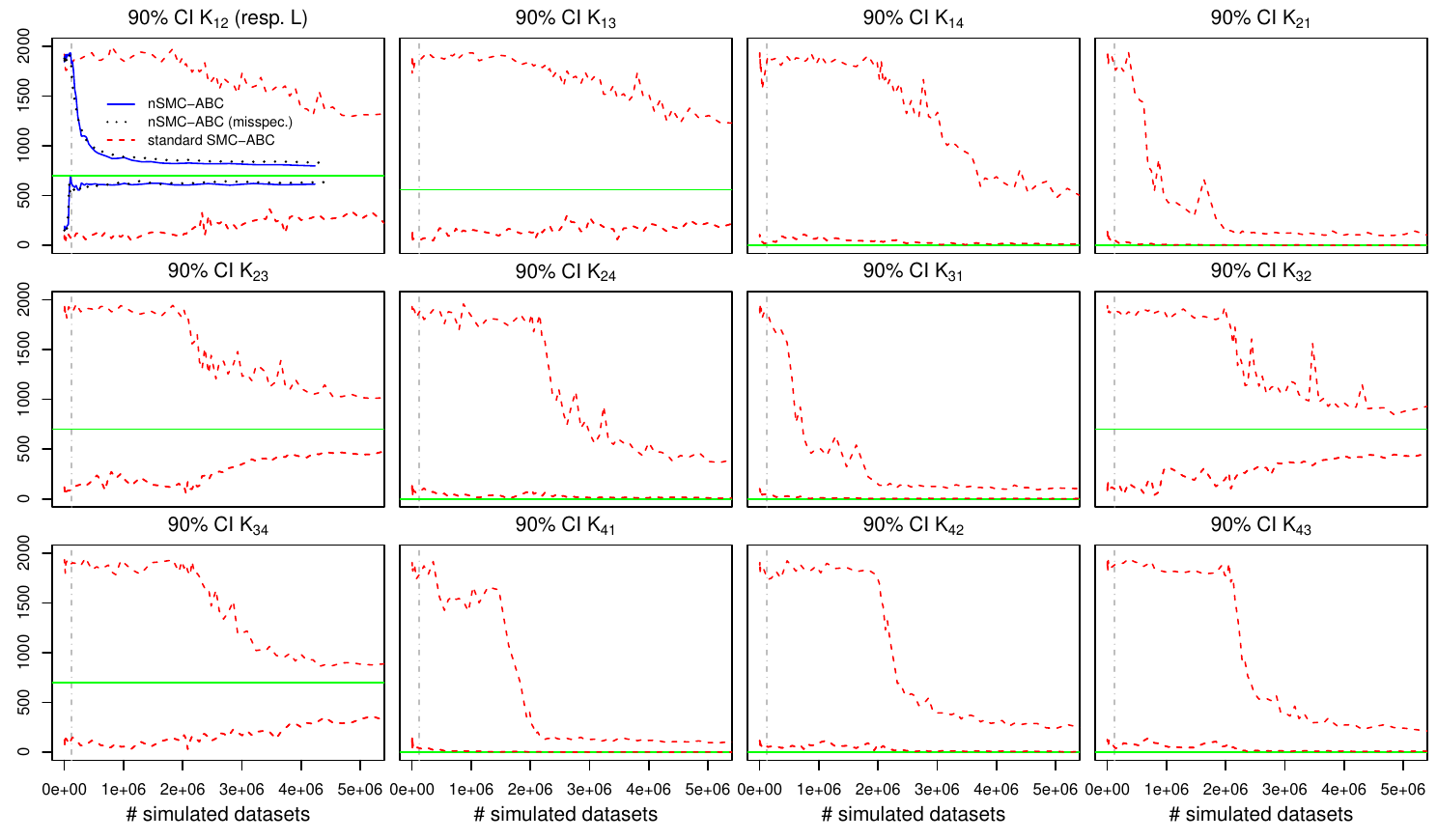}	
			\caption{\textbf{90\% CIs of the marginal posterior densities} of the $L$-parameter and the $K_{jk}$-parameters obtained from the nSMC-ABC Algorithm \ref{alg:SMC_SBP_ABC} (blue solid lines) and standard SMC-ABC (red dashed lines), respectively, as functions of the number of model simulations. The green horizontal lines indicate the true parameter values. The black dotted lines correspond to the inferential results obtained via nSMC-ABC under a misspecified model scenario. The  gray vertical dotted-dashed lines denote the number of simulations from which the correct network is inferred via nSMC-ABC. 
			}
			\label{fig:advantageBinaryKs}
		\end{centering}
	\end{figure}
	
	\begin{figure}[t]
		\begin{centering}
			\includegraphics[width=1.0\textwidth]{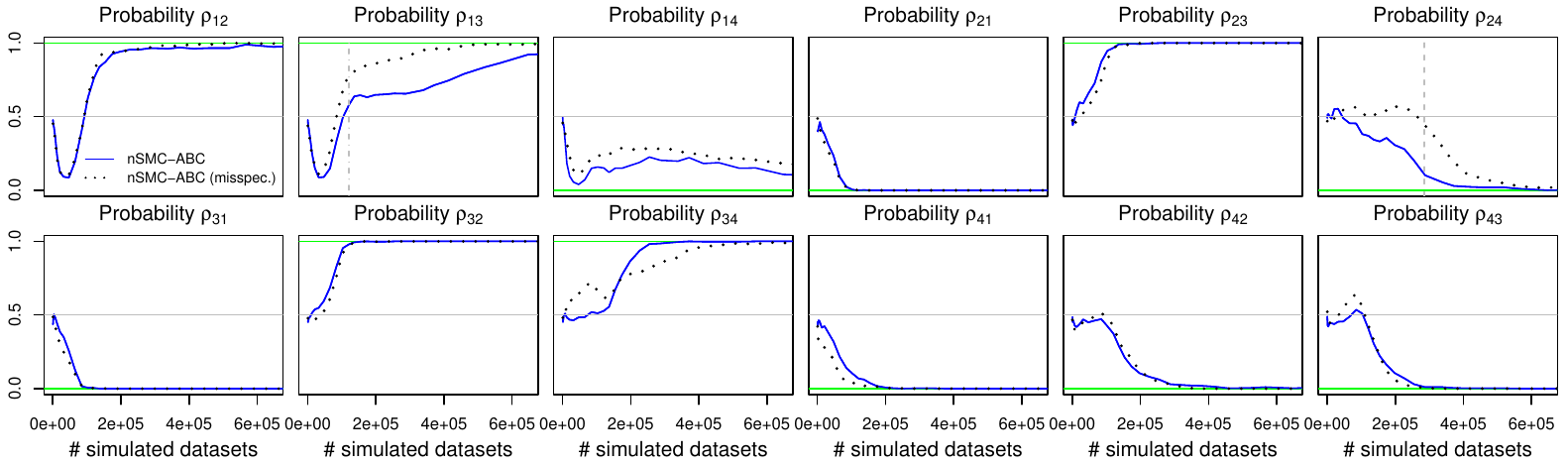}	
			\caption{\textbf{Marginal posterior means} of the $\rho_{jk}$-parameters obtained from the nSMC-ABC Algorithm \ref{alg:SMC_SBP_ABC}, as functions of the number of model simulations. The green horizontal lines indicate the true paramter values. The black dotted lines correspond to the inferential results obtained via nSMC-ABC under a misspecified model scenario. The  gray vertical dotted-dashed lines denote the number of simulations from which the correct network is inferred. All ABC posterior means are around $1/2$ at the beginning of the algorithm, as the $\rho_{jk}$ are sampled from Bernoulli priors. 
			}
			\label{fig:advantageBinaryRhos}
		\end{centering}
	\end{figure}
	
	\begin{figure}
		\begin{centering}
			\includegraphics[width=1.0\textwidth]{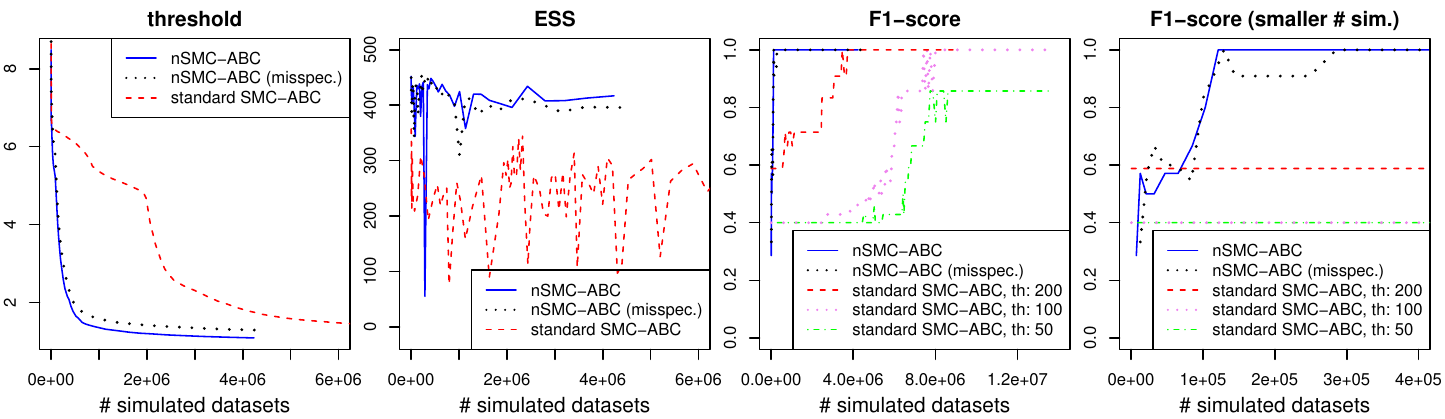}	
			\caption{\textbf{Distance-threshold $\delta$, ESS and F1-score} derived under the nSMC-ABC Algorithm \ref{alg:SMC_SBP_ABC} (blue solid lines) and the standard SMC-ABC algorithm (red dashed lines), as functions of the number of model simulations. The red dashed, violet dotted and green dotted-dashed lines for the F1-score correspond to the standard SMC-ABC algorithm using the network decision-threshold $200$, $100$ and $50$, respectively (to determine whether a connection is or not present with the standard SMC-ABC algorithm). The black dotted lines correspond to nSMC-ABC applied under a misspecified model scenario. 
			}
			\label{fig:advantageBinaryScores}
		\end{centering}
	\end{figure}
	
	In Figure \ref{fig:advantageBinaryAs}, we report the 90\% credible intervals (CIs) of the marginal posterior densities of the $A$-parameters derived under the proposed nSMC-ABC Algorithm \ref{alg:SMC_SBP_ABC} (blue solid lines) and the standard SMC-ABC algorithm (red dashed lines), as  functions of the number of datasets simulated from the model within the corresponding algorithm. We observe that the standard SMC-ABC algorithm needs far more model simulations (and thus a higher computational cost) to reach the desired posterior region, with the 90\% CI for $A_1$ not covering the true value. Instead, the proposed nSMC-ABC Algorithm \ref{alg:SMC_SBP_ABC} reaches the desired parameter regions much faster, requiring only a relatively small number of model simulations.
	
	Similarly, in Figure \ref{fig:advantageBinaryKs}, we report the 90\% CI of the marginal posterior densities of the $L$-parameter obtained under the nSMC-ABC Algorithm \ref{alg:SMC_SBP_ABC} (blue solid line) versus the 90\% CIs of the $K_{jk}$-parameters obtained under the standard SMC-ABC algorithm (red dashed lines). Again, we observe that the standard method requires far more model simulations than the proposed nSMC-ABC method. This becomes even more evident when looking at Figure \ref{fig:advantageBinaryRhos}, where we report the approximate posterior means of the binary parameters $\rho_{jk}$, inferred via nSMC-ABC, as a function of the number of model simulations. We observe that the correct network is already inferred for fewer than $2\cdot 10^5$ model simulations (see the corresponding gray vertical dotted-dashed lines in Figures \ref{fig:advantageBinaryKs} and \ref{fig:advantageBinaryRhos}), whereas at least around $3 \cdot 10^6$ simulations are required until one may deduce the correct network from the $K_{jk}$-posteriors visualized in Figure \ref{fig:advantageBinaryKs} via a suitable decision criterion. Specifically, for the standard SMC-ABC algorithm, we say that a connection from Population $j$ to population $k$ is present if the obtained posterior median of the parameter $K_{jk}$ is larger than some pre-fixed network decision-threshold. 
	
	These observations are also in agreement with the results shown in Figure \ref{fig:advantageBinaryScores}. There, we report the distance-threshold $\delta$, the effective sample size (ESS; a number in $[1, M]$ measuring  how many particles are ``relevant'' at each iteration), and the F1-score (a number in $[0,1]$ measuring the quality of the network estimation) as functions of the number of model simulations. For example, we observe that the threshold $\delta$ decreases much faster for the nSMC-ABC Algorithm \ref{alg:SMC_SBP_ABC} (blue solid line) than for the standard SMC-ABC (red dashed line). Moreover, the ESS is larger for the nSMC-ABC method. In addition, the nSMC-ABC Algorithm \ref{alg:SMC_SBP_ABC} yields an F1-score equal to $1$ (i.e., the entire network is correctly inferred) already for a very small number of simulations (between $10^5$ and $2\cdot 10^5)$, while the standard SMC-ABC algorithm requires around $5 \cdot 10^6$ model simulations. Note, however, that only when the threshold value for deciding whether a connection is present or not is large enough (here, $200$), the standard algorithm yields the correct network (for a sufficiently large number of model simulations). When such threshold is set to $100$, standard SMC-ABC requires even more model simulations for correct network inference (violet dotted lines), failing to infer the entire network correctly (green dotted-dashed lines) when the decision-threshold is reduced to $50$. This highlights the arbitrariness and impact of this coefficient, and the advantage of the proposed nSMC-ABC approach which does not rely on it. 
	
	To emphasize the advantage of the binary network indicators even more, we also investigate the nSMC-ABC algorithm under a misspecified model scenario. Specifically, we apply the nSMC-ABC Algorithm \ref{alg:SMC_SBP_ABC}, assuming the previously considered coupling structure $ K_{jk}:=c^{|j-k|-1}L$ (see \eqref{eq:coupling_structure}) to (simulated) reference data violating this network assumption. In particular, the reference data is obtained by generating all $K_{jk}$ from a continuous uniform distribution $\textrm{U}(500,800)$. The results are reported as black dotted lines in Figures \ref{fig:advantageBinaryAs}-\ref{fig:advantageBinaryScores}. Even though the model is now misspecified, we obtain almost equally good inferential results, with only the binary parameter $\rho_{24}$ requiring a few more model simulations to be correctly identified via nSMC-ABC. This also affects the F1-score (see the right panel of Figure \ref{fig:advantageBinaryScores}).
	
	We refer to Appendix \ref{App:CouplingStrengthStructure} for an investigation of different choices of the coupling strength structure \eqref{eq:mat_CouplingStrengthStructure}, all leading to the same inferred networks when applying the nSMC-ABC method to real EEG data.

	
	\subsection{\lowercase{n}SMC-ABC: Analysis of the perturbation kernel}
	\label{app:perturbationKernel}
	
	\begin{figure}
		\begin{centering}
			\includegraphics[width=1.0\textwidth]{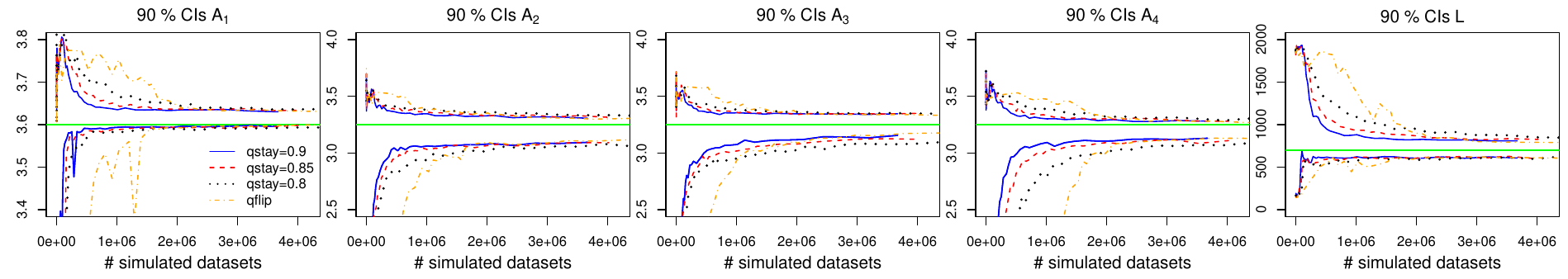}	
			\caption{\textbf{90\% CIs of the marginal posterior densities} of the continuous parameters (other than $c$) obtained from the nSMC-ABC Algorithm~\ref{alg:SMC_SBP_ABC} for $q_{\textrm{stay}}$ (see \eqref{kernel_d2}) equal to $0.9$ (blue solid lines), $0.85$ (red dashed lines) and $0.8$ (black dotted lines), as functions of the number of model simulations. The orange dotted-dashed lines correspond to the inferential results obtained via nSMC-ABC with an alternative perturbation kernel.}
			\label{fig:perturbationKernelAs}
		\end{centering}
	\end{figure}
	
	\begin{figure}
		\begin{centering}
			\includegraphics[width=1.0\textwidth]{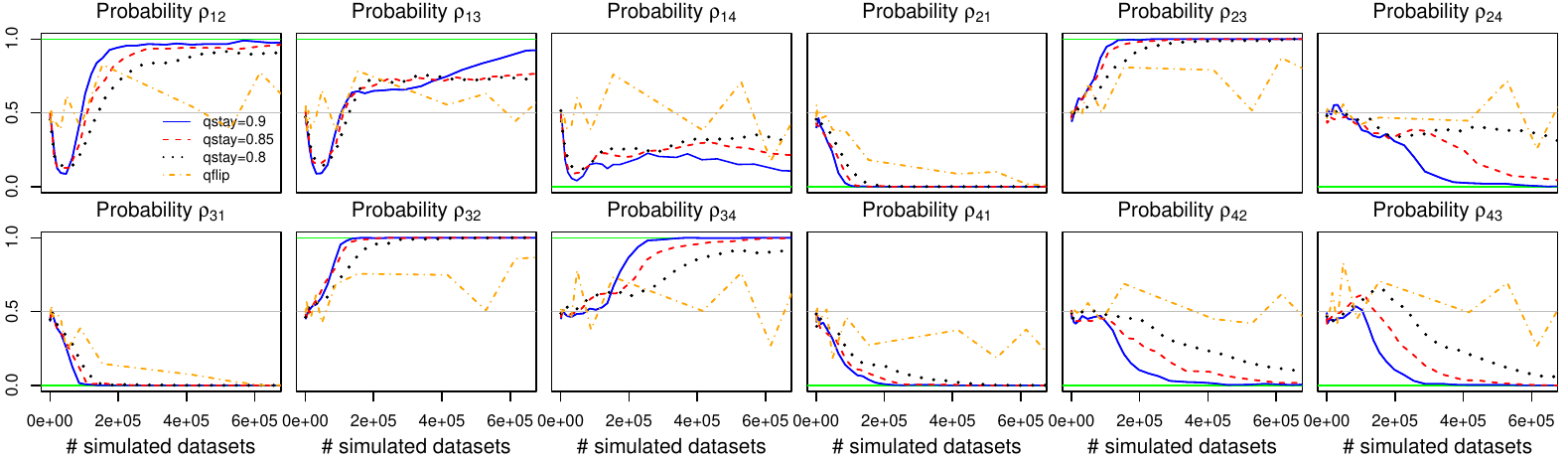}	
			\caption{\textbf{Marginal posterior means} of the $\rho_{jk}$-parameters obtained from the nSMC-ABC Algorithm \ref{alg:SMC_SBP_ABC} for $q_{\textrm{stay}}$ (see \eqref{kernel_d2}) equal to $0.9$ (blue solid lines), $0.85$ (red dashed lines) and $0.8$ (black dotted lines), as functions of the number of model simulations. The orange dotted-dashed lines correspond to the inferential results obtained via nSMC-ABC with an alternative perturbation kernel.}
			\label{fig:perturbationKernelRhos}
		\end{centering}
	\end{figure}
	
	\begin{figure}[t]
		\begin{centering}
			\includegraphics[width=1.0\textwidth]{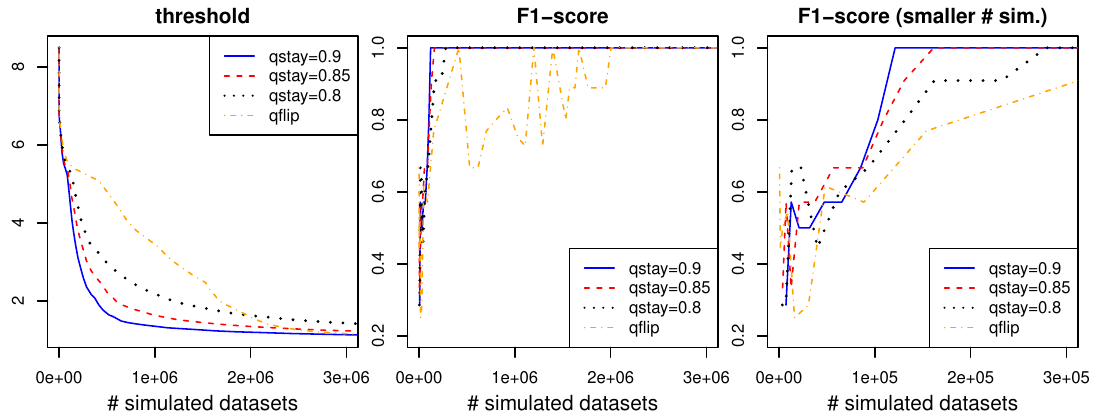}	
			\caption{\textbf{Distance-threshold $\delta$ and F1-score} derived under the nSMC-ABC Algorithm \ref{alg:SMC_SBP_ABC} for $q_{\textrm{stay}}$ (see \eqref{kernel_d2}) equal to $0.9$ (blue solid lines), $0.85$ (red dashed lines) and $0.8$ (black dotted lines), as functions of the number of model simulations. The orange dotted-dashed lines correspond to nSMC-ABC with an alternative perturbation kernel. 
			}
			\label{fig:perturbationKernelScores}
		\end{centering}
	\end{figure}
	
	In this section, we investigate more closely the proposed perturbation kernel of the nSMC-ABC Algorithm~\ref{alg:SMC_SBP_ABC}, focusing again on Setting 2 ``Partially connected network'', with similar results obtained for the other settings. 
	
	First, we consider the proposed Bernoulli-type kernel, focusing on different values for $q_{\textrm{stay}}$ (see \eqref{kernel_d2}), namely $q_{\textrm{stay}}=0.9$ (as in the main manuscript, blue solid lines), $q_{\textrm{stay}}=0.85$ (red dashed lines), and $q_{\textrm{stay}}=0.8$ (black dotted lines), all leading to similar results. However, the larger $q_{\textrm{stay}}$, the faster the desired marginal posterior region of the continuous parameters is reached (see Figure~\ref{fig:perturbationKernelAs}), the faster the binary parameters are correctly identified (see Figure~\ref{fig:perturbationKernelRhos} and the F1-scores in Figure~\ref{fig:perturbationKernelScores}), and the faster the distance-threshold $\delta$ decreases (see the left panel of Figure \ref{fig:perturbationKernelScores}). This indicates that the more we trust the information from the previous population, the better the algorithm performs. However, to prevent the binary particles from being caught in wrong parameter regions (which might be especially relevant in the first few iterations), we refrain from choosing $q_{\textrm{stay}}$ too close to $1$.
	
	Second, we consider an alternative to the the proposed Bernoulli-type kernel \eqref{kernel_d2}, which preserves some dependency between the continuous and binary parameters, as well as within the binary parameters. This alternative kernel first samples a full particle (both continuous and binary) and then perturbs it, as commonly done in standard SMC-ABC. For the perturbation, it uses again the multivariate Gaussian kernel for the continuous parameters and it flips some of the binary parameters at random. In particular, the $u$-th sampled binary parameter $\theta_b^u$ is perturbed to $1-\theta_b^u$ with probability
	\begin{equation}\label{eq:flip}
		q^u_{\textrm{flip}}=4 \hat{v}_{u,r},
	\end{equation}
	where $\hat{v}_{u,r}$ is the population variance obtained from the previous population of that binary parameter at iteration $r-1$. The larger the variance  in the previous population of a binary parameter, the larger the probability that the sampled value is flipped (i.e., the less we trust the sampled value). This is in agreement with the perturbation of the continuous parameters, in the sense that the larger the variance in the previous population, the ``stronger'' the perturbation~is.
	
	\begin{remark}
		We use the population variance in \eqref{eq:flip}, because, in contrast to the sample variance, it is always between $0$ and $1/4$, guaranteeing $q_{\textrm{flip}}$ to be a probability value within~$[0,1]$.
	\end{remark}
	
	We apply the nSMC-ABC method based on this alternative perturbation kernel to the same observed data as before, and report the results as orange dotted-dashed lines in Figures \ref{fig:perturbationKernelAs}-\ref{fig:perturbationKernelScores}. We observe that also this version of the algorithm leads to similar marginal posteriors as nSMC-ABC using the proposed perturbation kernel, requiring significantly more model simulations to reach the desired parameter regions though. This becomes evident when looking at Figure \ref{fig:perturbationKernelAs} and \ref{fig:perturbationKernelRhos}, where we report the $90$\% CIs of the marginal posterior densities of the continuous parameters and the marginal posterior means of the $\rho_{jk}$-parameters, respectively, as functions of the number of model simulations. Moreover, the distance-threshold $\delta$ decreases slower in the alternative nSMC-ABC algorithm (cf. left panel of Figure \ref{fig:perturbationKernelScores}), which requires more model simulations (and thus more computational time) to correctly infer the network  
	(cf. the F1-scores in the middle and right panel of Figure \ref{fig:perturbationKernelScores}). This indicates that the proposed nSMC-ABC method yields satisfactory inferential results for different types of perturbation kernels, while benefiting from an independent treatment of the continuous and binary parameters, as proposed in Algorithm \ref{alg:SMC_SBP_ABC}.
	
	\begin{remark}
		The optimal local covariance matrix (\texttt{olcm}) Gaussian proposal sampler introduced in \cite{Filippi2013} is an alternative to the Gaussian perturbation kernel considered in the manuscript. As previously  observed (e.g., in \cite{Filippi2013,PicchiniTamborrino2022,Samson2025}), we find that \texttt{olcm} requires fewer model simulations to reach the desired posterior regions (results not shown). However, this advantage comes at the price of more involved proposal mean and covariance computations. 
	\end{remark}
	
	
	\section{Analysis of EEG recordings}
	\label{app:otherEEGrecordings}
	In this section, we deepen and extend the real data application of the proposed nSMC-ABC method. In Section \ref{App:CouplingStrengthStructure}, we illustrate that the inferential results presented in Section \ref{sec:6} of the main manuscript are robust under different choices of the underlying coupling strength structure. In Section \ref{app:otherEEGpatient}, we apply the proposed method to two further EEG recordings from another patient.  
	

	\subsection{Impact of the choice of the coupling strength structure}
	\label{App:CouplingStrengthStructure}

	Here, we consider again the EEG recording {\tt data1} visualized in Figure \ref{fig:EEG_data_chb01}a and repeat the inference presented in Section~\ref{sec:6} of the main manuscript (based on the coupling strength structure $K$ \eqref{eq:mat_CouplingStrengthStructure}) under two further choices of the coupling strength structure, namely
	
	\begin{equation}\label{eq:Ks}
		\bar{K}=\begin{pmatrix}
			- & L & L & L  \\
			L & -  & L & L  \\
			L & L & - & L \\
			L & L & L & -
		\end{pmatrix}, \qquad \widetilde{K}=\begin{pmatrix}
			- & L_1 & L_2 & L_2  \\
			L_1 & -  & L_2 & L_2  \\
			L_2 & L_2 & - & L_1 \\
			L_2 & L_2 & L_1 & -
		\end{pmatrix}.
	\end{equation}
	
	The first structure $\bar{K}$ assumes an identical and unknown coupling strength $L>0$ among all populations, fixing the scale parameter $c=1$ in \eqref{eq:mat_CouplingStrengthStructure}. The second structure $\widetilde{K}$ considers unknown coupling strength parameters $L_1,L_2>0$ for within and across hemisphere connections, respectively.
	
	\begin{figure}
		\begin{centering}
			\includegraphics[width=1.0\textwidth]{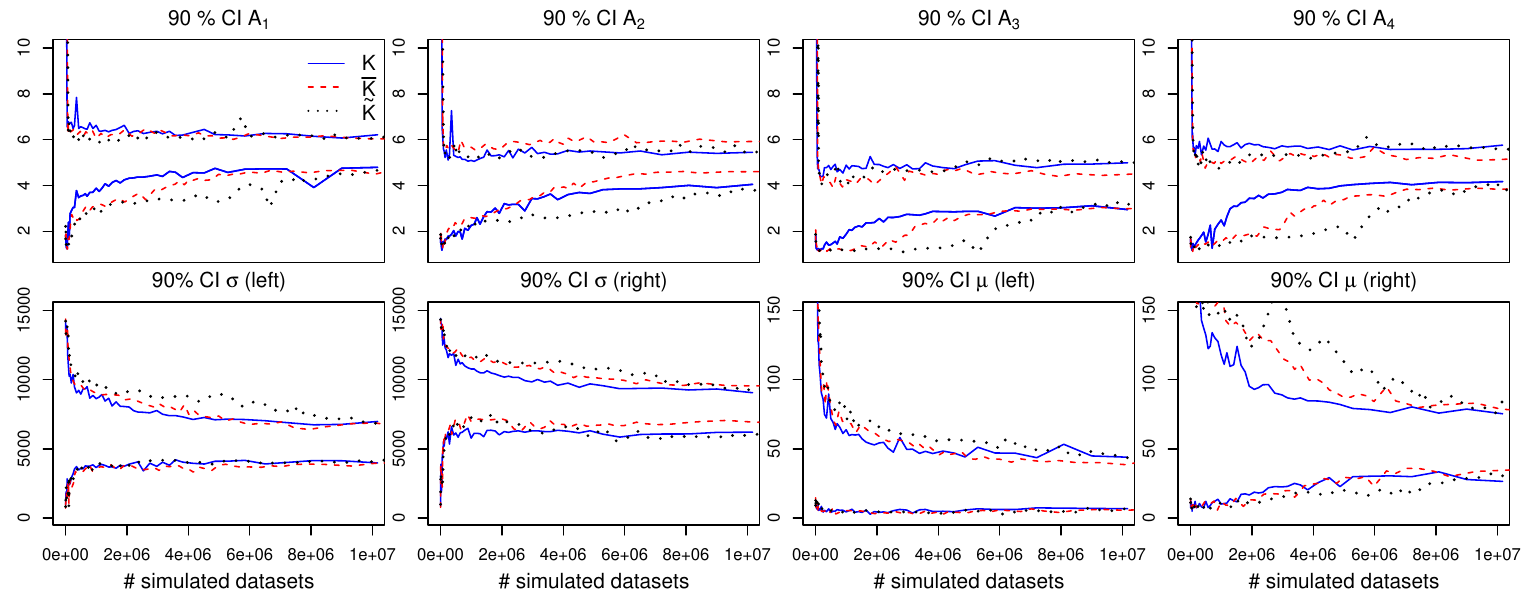}	
			\caption{\textbf{90\% CIs of the marginal posterior densities} of the continuous model parameters obtained via the nSMC-ABC Algorithm \ref{alg:SMC_SBP_ABC} from the during seizure period of {\tt data1} (cf. Figure \ref{fig:EEG_data_chb01}a), as functions of the number of model simulations. The solid blue, dashed red and dotted black lines correspond to the coupling strength structures $K$ \eqref{eq:mat_CouplingStrengthStructure}, $\bar{K}$ and $\widetilde{K}$ \eqref{eq:Ks}, respectively. Similar results are obtained from the before seizure period of~{\tt data1}. 
			}
			\label{fig:CSstructure_cont_during}
		\end{centering}
	\end{figure}

	We apply the proposed nSMC-ABC Algorithm \ref{alg:SMC_SBP_ABC} to the before and during seizure periods of {\tt data1} (cf. Figure \ref{fig:EEG_data_chb01}a) under the coupling strength structures $K$ \eqref{eq:mat_CouplingStrengthStructure}, $\bar{K}$ and $\widetilde{K}$ \eqref{eq:Ks}, respectively, performing thus six inferential tasks. The results for $K$ \eqref{eq:mat_CouplingStrengthStructure} correspond to those presented in Section \ref{sec:6} of the main manuscript.
	In Figure \ref{fig:CSstructure_cont_during}, we report the $90 \%$ CIs of the marginal posterior densities of the continuous model parameters for the during seizure period under the coupling strength structure $K$ \eqref{eq:mat_CouplingStrengthStructure} (solid blue lines), $\bar{K}$ \eqref{eq:Ks} (dashed red lines) and $\widetilde{K}$ \eqref{eq:Ks} (dotted black lines), respectively, as functions of the number of model simulations. Similar results are obtained for the before seizure period. It can be observed that all three different coupling strength structures lead to very similar inferential results for the continuous parameters, the algorithm converging slightly faster to the final posterior regions under $K$~\eqref{eq:mat_CouplingStrengthStructure}. This is particularly noticeable, e.g., in the right panels.
	
\begin{figure}[t]
		\begin{centering}
			\includegraphics[width=1.0\textwidth]{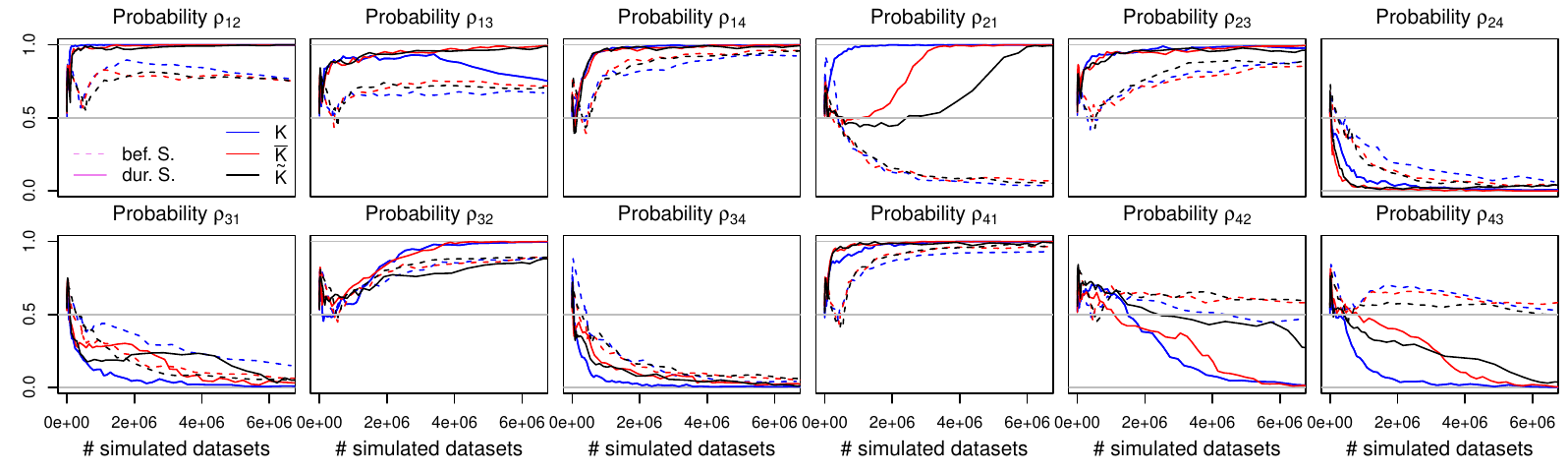}	
			\caption{\textbf{Marginal posterior means} of the $\rho_{jk}$-parameters obtained via the nSMC-ABC Algorithm \ref{alg:SMC_SBP_ABC} from the before (dashed lines) and during (solid lines) seizure period of {\tt data1} (cf. Figure \ref{fig:EEG_data_chb01}a), as functions of the number of model simulations. The corresponding blue, red and black lines are obtained under the coupling strength structures $K$~\eqref{eq:mat_CouplingStrengthStructure}, $\bar{K}$ and $\widetilde{K}$ \eqref{eq:Ks}, respectively.}
			\label{fig:CSStructure_rhos}
		\end{centering}
	\end{figure}

	These findings are confirmed by looking at Figure \ref{fig:CSStructure_rhos}, where we report the corresponding marginal posterior means of the binary coupling direction parameters $\rho_{jk}$, $j,k=1,\ldots,4$, $j\neq k$, as functions of the number of model simulations, for the before (dashed lines) and during (solid lines) seizure periods of {\tt data1} (cf. Figure \ref{fig:EEG_data_chb01}a). The respective blue, red and black lines correspond to $K$ \eqref{eq:mat_CouplingStrengthStructure}, $\bar{K}$ and $\widetilde{K}$ \eqref{eq:Ks}, respectively. Also for the binary parameters, the inferential results are similar under the three different coupling structures, leading to the same inferred networks. Moreover, for some of the network parameters, we observe a faster convergence of the algorithm under $K$ \eqref{eq:mat_CouplingStrengthStructure} (to 0/1) in the during seizure setting. This is particularly evident for $\rho_{21}$, but also, for example, for $\rho_{31}$, $\rho_{34}$ and $\rho_{43}$. Note also that the results for $\rho_{42}$ and $\rho_{43}$ under the before seizure scenario confirm the low evidence that we have for those connections to be present or not, as indicated with the dotted orange lines in Figure \ref{fig:ABC_results_EEG_net}a (left network).
	
	\begin{figure}
		\begin{centering}
			\includegraphics[width=0.85\textwidth]{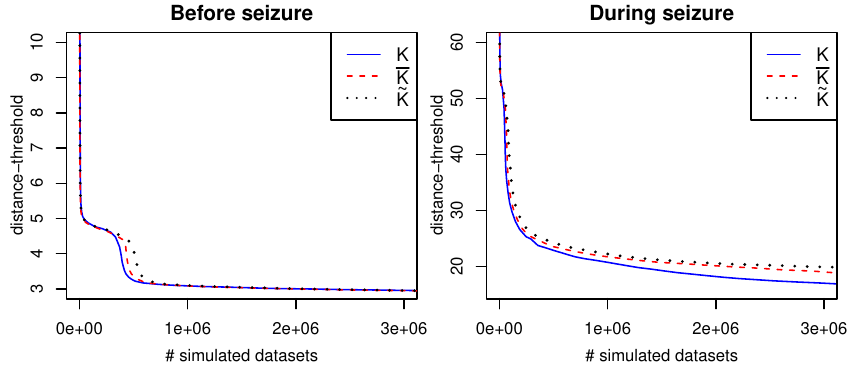}	
			\caption{\textbf{Distance-threshold $\delta$} obtained via the nSMC-ABC Algorithm \ref{alg:SMC_SBP_ABC} from the before (left panel) and during (right panel) seizure period of {\tt data1} (cf. Figure \ref{fig:EEG_data_chb01}a), as functions of the number of model simulations. The corresponding solid blue, dashed red and dotted black lines are obtained under the coupling strength structures $K$~\eqref{eq:mat_CouplingStrengthStructure}, $\bar{K}$ and $\widetilde{K}$ \eqref{eq:Ks}, respectively.
			}
			\label{fig:CSStructure_budget_vs_threshold}
		\end{centering}
	\end{figure}

	The slightly faster convergence of the nSMC-ABC Algorithm \ref{alg:SMC_SBP_ABC} under the coupling strength structure $K$ \eqref{eq:mat_CouplingStrengthStructure} is also supported by Figure \ref{fig:CSStructure_budget_vs_threshold}, where we report the distance-threshold $\delta$ (which decreases along iterations of the algorithm), as function of the number of model simulations, for the before (left panel) and during (right panel) seizure periods of {\tt data1} (cf. Figure \ref{fig:EEG_data_chb01}a), under the coupling strength structure $K$ \eqref{eq:mat_CouplingStrengthStructure} (solid blue lines), $\bar{K}$ \eqref{eq:Ks} (dashed red lines) and $\widetilde{K}$ \eqref{eq:Ks} (dotted black lines), respectively. It can be observed that the threshold for the distances within Algorithm~\ref{alg:SMC_SBP_ABC} decreases slightly faster under $K$ \eqref{eq:mat_CouplingStrengthStructure} for both scenarios, suggesting a faster convergence of the algorithm towards parameter regions which yield smaller distances. This is remarkable, considering the fact that under $\bar{K}$ \eqref{eq:Ks} one continuous parameter less has to be inferred.

	Overall, these results suggest both robustness of the inference for real data under different coupling strength structures, and plausibility of the assumed coupling strength structure \eqref{eq:mat_CouplingStrengthStructure}~$\text{(cf.~\eqref{eq:coupling_structure})}$.
	
	
	\subsection{Inference from another patient's EEG recordings}
	\label{app:otherEEGpatient}

\begin{figure}[H]
		\centering
		\subfigure[{\tt data1}: before und during seizure]{\includegraphics[width=0.85\textwidth]{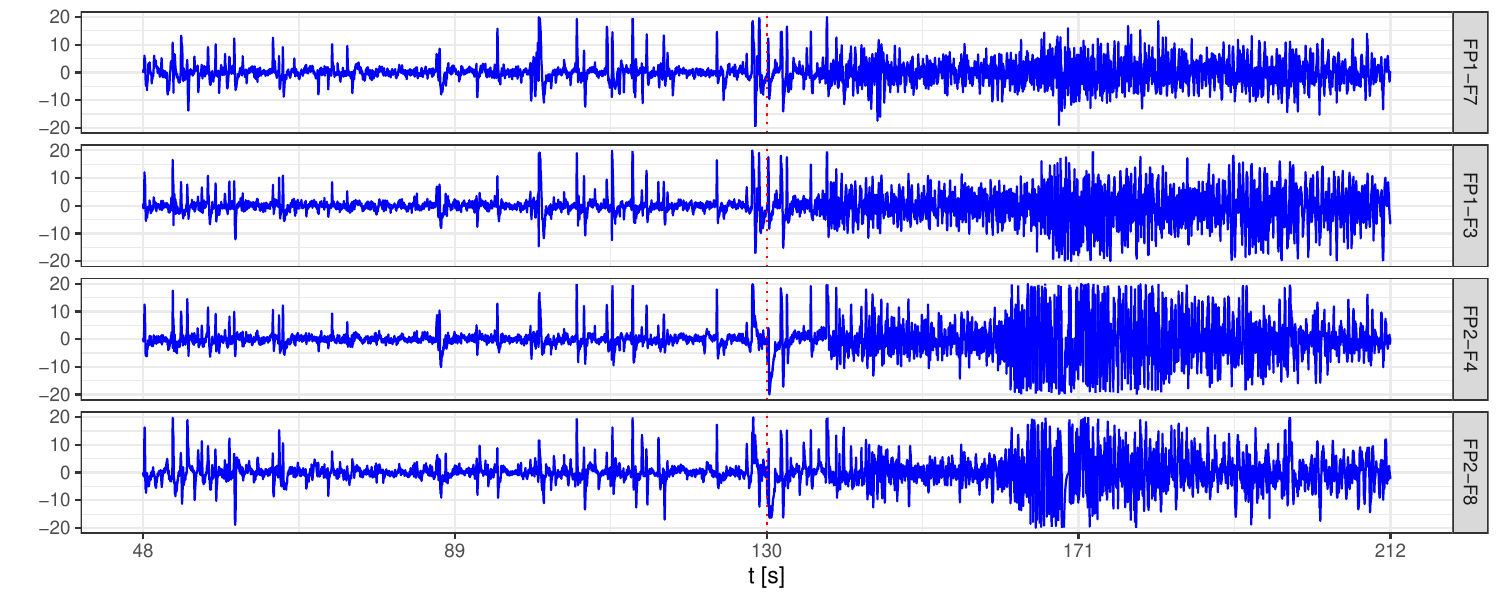}}
		\subfigure[{\tt data2}: before und during seizure]{\includegraphics[width=0.85\textwidth]{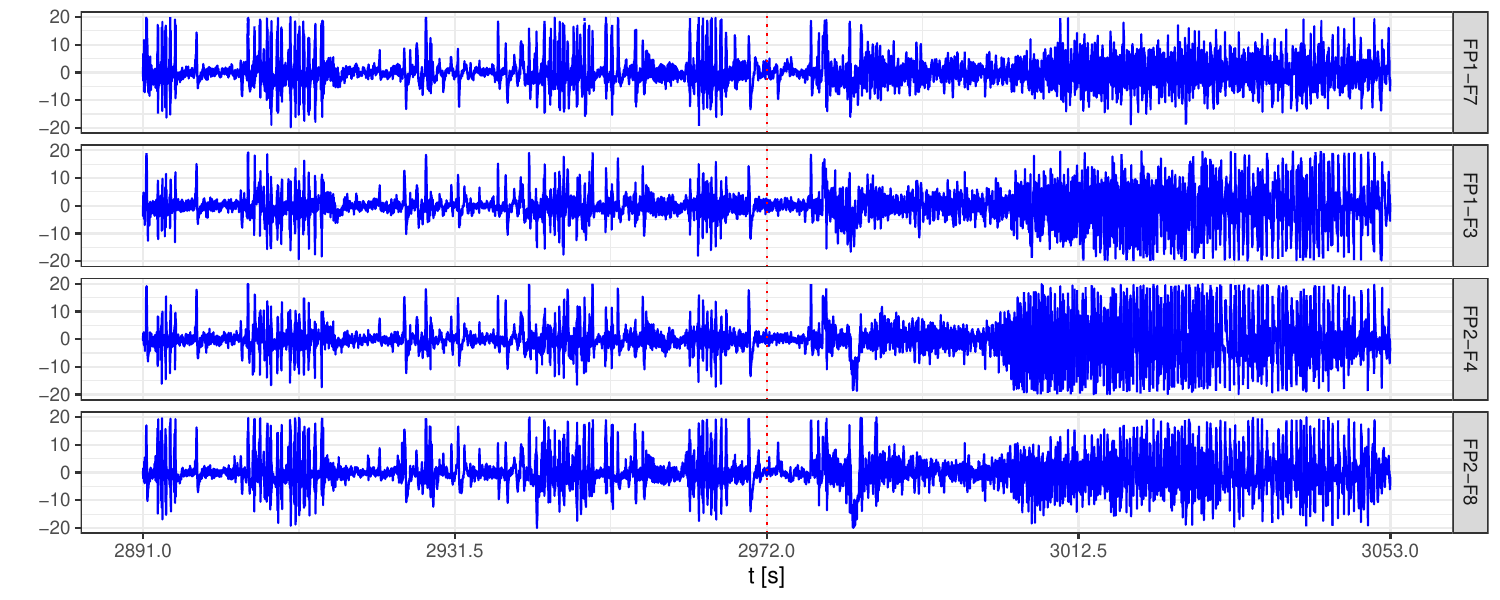}} 
		\vspace{-0.2cm}
		\caption{{\bf Two EEG recordings of an $11$ year old male patient.} The intervals $[130,112]s$ in {\tt data1} (a) and  $[2972,3053]s$ in {\tt data2} (b) (measurements to the right of the vertical dotted red lines) have been classified as a seizure in \cite{Shoeb2009}.} 
		\label{fig:EEG_data_chb02}
	\end{figure}

	In this section, we investigate two EEG recordings of the second pediatric subject from the \textit{CHB-MIT Scalp EEG Database}, an $11$ year old male patient (see Section \ref{sec:6:1} of the main manuscript for the data details). The two investigated datasets are recorded over time intervals of length $164$ and $162$ seconds, respectively. They are available in the edf-files \texttt{chb02\_16} and \texttt{chb02\_16+}, and denoted by \texttt{data1} and \texttt{data2} in the following. A visualization of these datasets is provided in Figure \ref{fig:EEG_data_chb02}, where the dotted red lines separate the data  into the respective period before and during seizure (see the seizure classification in \cite{Shoeb2009}). As in Section \ref{sec:6:2} of the main manuscript, we denote the channels FP1-F7, FP1-F3, FP2-F4, and FP2-F8 by Population $1$, $2$, $3$, and $4$, respectively, and apply the proposed nSMC-ABC Algorithm \ref{alg:SMC_SBP_ABC} (under the coupling strength structure $K$ \eqref{eq:mat_CouplingStrengthStructure}) to fit the stochastic multi-population JR-NMM to these four EEG segments.
	
\begin{figure}[H]
		\begin{centering}
			\includegraphics[width=1.0\textwidth]{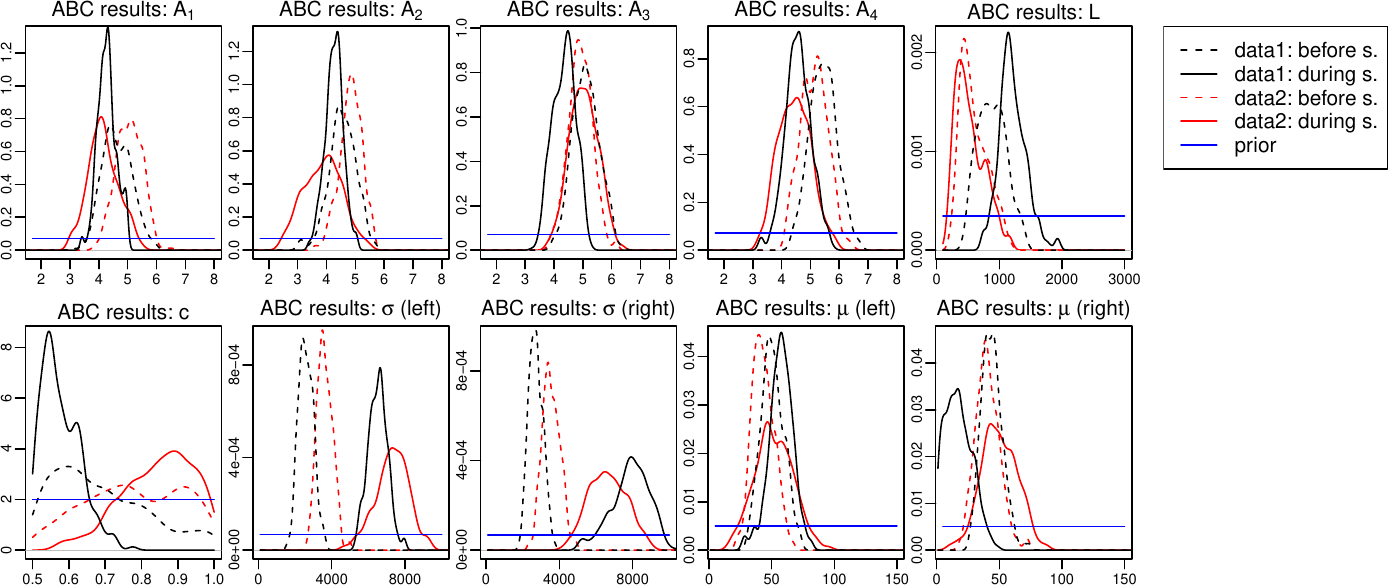}	
			\caption{\textbf{nSMC-ABC marginal  
					posterior densities} of the continuous parameters before (dashed lines) and during (solid lines) seizure ({\tt data1}: black, {\tt data2}: red). The horizontal blue lines are the respective uniform prior densities.}
			\label{fig:ABC_results_EEG_chb02_cont}
		\end{centering}
	\end{figure}

	In Figure \ref{fig:ABC_results_EEG_chb02_cont}, we report the marginal posterior densities of the continuous parameters of the first (black lines) and the second (red lines) dataset (before seizure: dashed lines, during seizure: solid lines). The uniform priors are indicated by the horizontal blue lines. Interestingly, while for the first patient we observed larger individual activation parameters $A_k$, $k=1,\ldots,4$, during seizure (cf. Figure \ref{fig:ABC_results_EEG_cont}), they are now smaller for the second patient. 
	This is  consistent for both datasets, the marginal posterior supports during seizure (solid black lines for {\tt data1} and red lines for {\tt data2}) being shifted to the left (i.e., to smaller parameter regions) compared to those before seizure (dashed black lines for {\tt data1} and red lines for {\tt data2}). Both patients (and both individual datasets) have larger noise intensity parameters $\sigma_L$ and $\sigma_R$ in both the left and the right brain hemisphere during seizure, the marginal posterior supports being clearly shifted to the right. In general, the second patient's results for the continuous model parameter are similar across the two datasets, with unimodal shaped posteriors. Moreover, while for the first subject we only obtained clear posterior estimates for the coupling strength parameter $L$ during seizure, for the second patient all posteriors for $L$ have a clear peak, with a larger coupling strength obtained~for~{\tt data1}.

	\setlength{\tabcolsep}{1pt}
	\begin{table}[H]
		{  
			\caption{nSMC-ABC network estimates of ${\rho}_{jk}$ obtained as marginal posterior medians (marginal posterior means in parentheses) for the 
				before and during seizure periods of {\tt data1} and {\tt data2}.}
			\vspace{-0.5cm}
			\label{table:ABC_results_rho_eeg_chb02}
			\begin{center}
				\scalebox{0.8}{
					\begin{tabular}{l|llllllllllll}
						\hline 
						EEG data & $\hat\rho_{12}$ & $\hat\rho_{13}$ & $\hat\rho_{14}$ & $\hat\rho_{21}$ & $\hat\rho_{23}$ & $\hat\rho_{24}$ & $\hat\rho_{31}$ & $\hat\rho_{32}$ & $\hat\rho_{34}$ & $\hat\rho_{41}$ & $\hat\rho_{42}$ & $\hat\rho_{43}$ \\  
						\hline
						{\tt data1}: b.s. & 0 (0.032) & 1 (0.926) & 0 (0.346) & 1 (0.994) & 1 (0.988) & 0 (0.254) & 1 (0.888) & 1 (0.986) & 1 (0.986) & 1 (0.752) & 1 (0.996) & 0 (0) \\	
						{\tt data1}: d.s. & 1 (1) & 1 (1) & 1 (0.972) & 0 (0) & 0 (0.004) & 0 (0) & 1 (0.986) & 0 (0) & 1 (1) & 0 (0.016) & 0 (0.056) & 1 (1) \\	
						\hline
						{\tt data2}: b.s. & 0 (0.164) & 1 (0.972) & 1 (0.754) & 1 (0.820) & 1 (0.994) & 1 (0.866) & 1 (0.662) & 1 (0.976) & 1 (0.924) & 1 (0.704) & 1 (0.954) & 1 (0.130)  \\	
						{\tt data2}: d.s. & 1 (1) & 0 (0.086) & 1 (1) & 1 (0.974) & 1 (0.980) & 0 (0.014) & 0 (0.002) & 1 (1) & 0 (0) & 1 (0.834) & 0 (0.162) & 1 (0.994) \\	
						\hline
				\end{tabular}}
		\end{center}}
	\vspace{-0.5cm}
	\end{table}  
	
	\begin{figure}[H]
		\centering
		\subfigure[Inferred network for {\tt data1}: before (left) and during (right) seizure]
		{\includegraphics[width=0.95\textwidth]{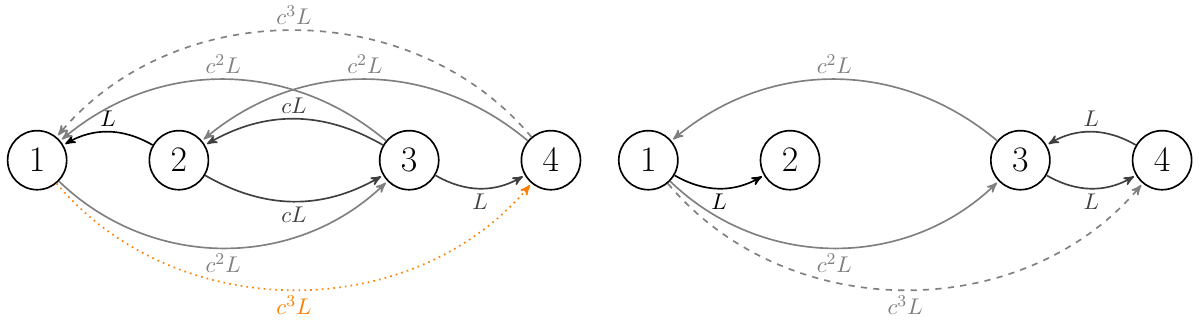}}
		\subfigure[Inferred network for {\tt data2}: before (left) and during (right) seizure]
		{\includegraphics[width=0.95\textwidth]{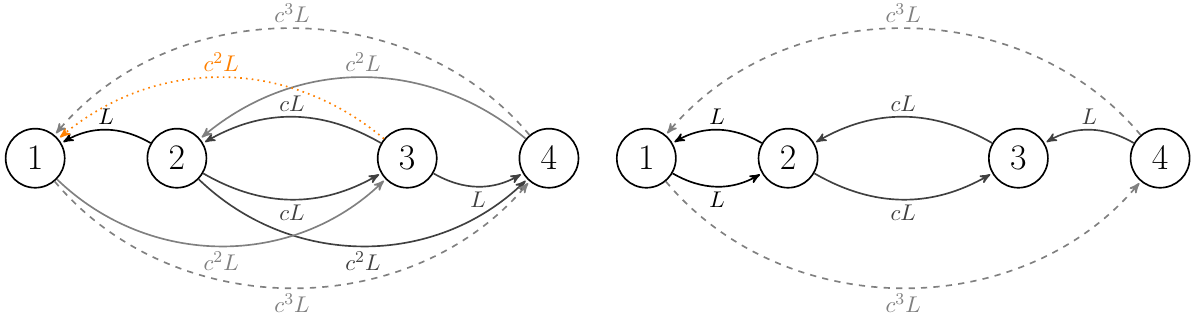}}
		\caption{{\bf Inferred networks from EEG recordings.} Estimated networks from two recordings with $N=4$ channels of an 11 year old male patient before and during epileptic seizure. Populations 1, 2, 3, and 4 refer to channels FP1-F7, FP1-F3, FP2-F4, and FP2-F8, respectively. The dotted orange connections are estimated with a posterior mean within $[1/3,2/3]$  
			(cf. Table \ref{table:ABC_results_rho_eeg_chb02}).  
		}
		\label{fig:ABC_results_EEG_net_chb02}
	\end{figure}

	In Table \ref{table:ABC_results_rho_eeg_chb02}, we report the network estimates of the binary $\rho_{jk}$-parameters and the corresponding marginal posterior means in parentheses. Similarly to the first subject, all posterior means for the during seizure scenario are either close to $1$ or $0$, clearly indicating whether a connection is present or not. For the before seizure scenarios, only the posterior means of $\rho_{14}$ for {\tt data1} and $\rho_{31}$ for {\tt data2} lie inside $[1/3,2/3]$, the possible unclear connections marked as dotted orange lines in Figure~\ref{fig:ABC_results_EEG_net_chb02}, where we visualize the inferred networks. As for the first patient, we observe some similarities across {\tt data1} and {\tt data2}. For example, in both recordings, we observe a very clear activation of $\rho_{12}$ in the left brain hemisphere  and of $\rho_{43}$ in the right brain hemisphere during seizure, connections which were both not present before seizure. Moreover, there is less connectivity between the two hemispheres during seizure, both from left to right and from right to left.
	
\begin{figure}[H]
		\begin{centering}
			\includegraphics[width=1.0\textwidth]{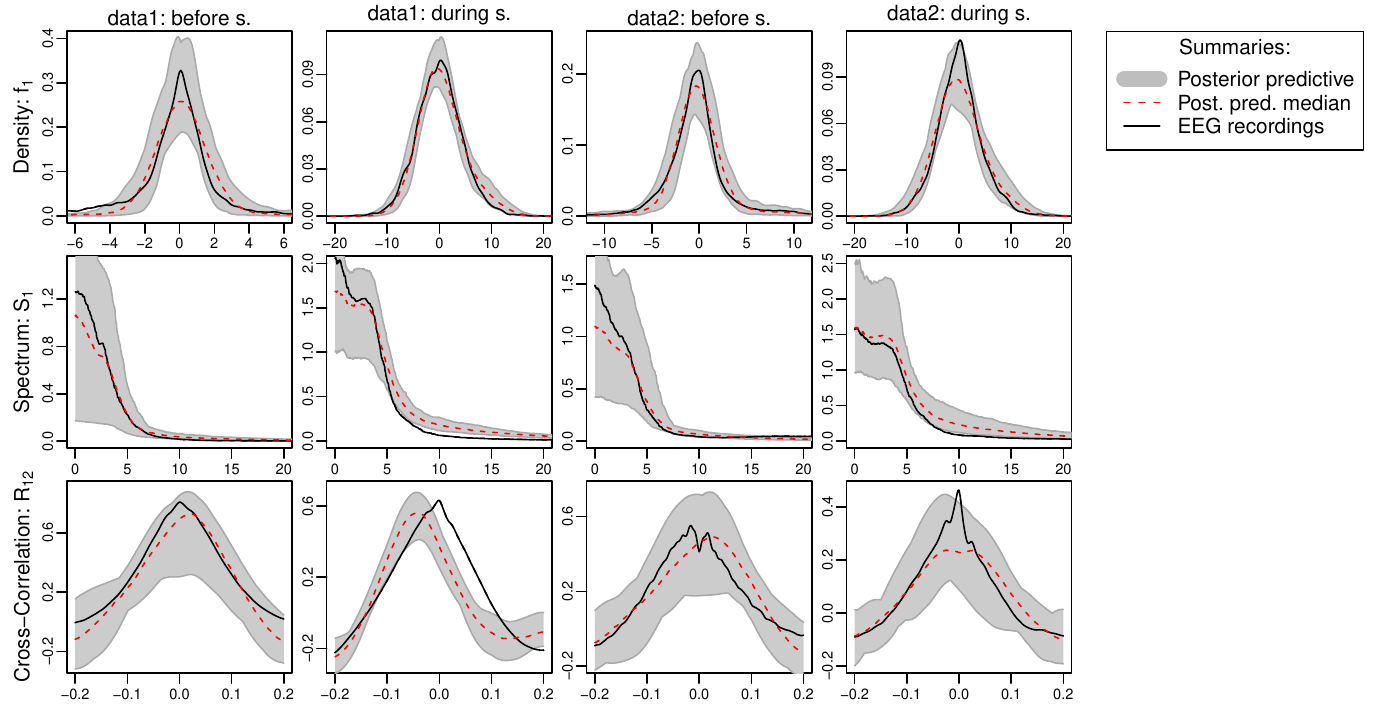}	
			\caption{\textbf{Summaries} $f_1$, $S_1$ and $R_{12}$ (cf. \eqref{eq:ABC_summaries}) of the EEG recordings (solid black lines) compared to summaries derived from synthetic datasets generated with $100$ kept posterior samples (gray bands and dashed red lines for their medians).}
			\label{fig:fittedSummaries_chb02}
		\end{centering}
	\end{figure}

	Finally, the summary statistics \eqref{eq:ABC_summaries} computed from the four EEG segments (solid black lines) are compared with those obtained from the posterior predictive (gray areas) in Figure~\ref{fig:fittedSummaries_chb02}, where the dashed red lines represent the medians of the posterior predictive bands. Again, only the subset $\{ f_1,S_1,R_{12} \}$ of the summaries is shown, noting that a similar fit is obtained for the other summary functions. As for the first subject, the match of the observed and posterior predicted summaries is generally very good.

\end{document}